%% file: src/00_Author_tex.tex
\documentclass{ptephy_v2}

\preprintnumber{XXXX-XXXX} 
\usepackage{hyperref}
\hypersetup{
 colorlinks = true,
 allcolors = blue,
 filecolor = magenta, 
 urlcolor = cyan,
 pdfborder = {0 0 0}
}

\usepackage{graphics} 
\usepackage{subcaption}
\usepackage{float}
\usepackage{color}
\usepackage{array}
\usepackage{multirow}

\begin{document}

\title{Performance evaluation of Luxium Solutions BCF-XL wavelength-shifting fibers} 


\author{\name{\fname{Tatsuki}~\surname{Yamazumi}}{*}}
\author{\name{\fname{Yota}~\surname{Endo}}{}}
\author{\name{\fname{Shoma}~\surname{Kodama}}{}}
\author{\name{\fname{Kota}~\surname{Nakagiri}}{}\thanks{Present address: Kamioka Observatry, Institute for Cosmic Ray Research, University of Tokyo, Kamioka, Gifu 506-1205, Japan}}
\author{\name{\fname{Yasuhiro}~\surname{Nakajima}}{}}
\author{\name{\fname{Minoru}~\surname{Sekiyama}}{}}
\author{\name{\fname{Masashi}~\surname{Yokoyama}}{}}
\affil{Department of Physics, Graduate School of Science, The University of Tokyo, 7-3-1 Hongo, Bunkyo-ku, Tokyo 113-0033, Japan \email{yamazumi@hep.phys.s.u-tokyo.ac.jp}}

\begin{abstract}%
We evaluate the performance of single-clad wavelength-shifting fibers newly developed by Luxium Solutions, BCF-92XL, BCF-9929AXL, and BCF-9995XL and compare them with the multi-clad Kuraray Y-11 fiber.
The BCF-XL fibers exhibit faster decay times (92XL: $2.10\pm0.01$~ns, 9929AXL: $2.10\pm0.02$~ns, 9995XL: $2.41\pm0.03$~ns) than Y-11 ($7.44\pm0.06$~ns). 
The attenuation lengths are comparable to that of Y-11 within the measurement range up to 3.2~m. 
When coupled to an EJ-204 plastic scintillator, the BCF-XL fibers achieve superior time resolution while maintaining light yields comparable to those expected for a single-clad Y-11 fiber.
\end{abstract}

\subjectindex{H10, H15}

\maketitle

\input{src/01_intro}
\input{src/02_laser}
\input{src/03_electronbeam}

\input{src/04_conclusion}

\section*{Acknowledgments}
We thank Luxium Solutions for providing samples of BCF-XL series WLS fibers. 
We gratefully acknowledge the support and assistance of all personnel involved in the PF-AR Test Beam Line of the Instrumentation Technology Development Center at KEK.
This work was supported by JSPS KAKENHI Grant Numbers JP20H00149, JP22H04943, JP24K00637, JP25H00004.


%


\let\doi\relax


\end{document}

%% file: src/01_intro.tex
\section{Introduction}

Scintillator-based particle detectors coupled with wavelength-shifting (WLS) fibers are widely employed in high-energy physics experiments, offering a practical solution for constructing large-scale detectors. Despite its widespread use, the time resolution of such systems is often limited by the decay time of WLS fibers, which is typically slower than the intrinsic response of the scintillator. Improving the timing properties of WLS fibers is therefore an important subject in the development of fast scintillator-based detector systems.


The Kuraray Y-11 fiber~\cite{y11_site} has served as the established standard in scintillator-based detectors due to its high light yield and long attenuation length. With the increasing demand for fast timing performance, Kuraray has introduced the YS series~\cite{2}, which offers a reduced decay time. Their optical and timing properties have been evaluated in previous studies, demonstrating clear improvements over Y-11~\cite{y11, kodamasan}. Recently, Luxium Solutions has developed a new type of WLS fiber, the BCF-XL series~\cite{1}. The key characteristics reported in the manufacturers' data sheets are summarized in Table~\ref{char_fibers}.

In this paper, we present a characterization of the Luxium BCF-92XL, BCF-9929AXL, and BCF-9995XL fibers and compare their performance with the Kuraray Y-11 fiber. All fibers are 1~mm in diameter. The BCF-XL series are single-clad, while the Y-11 fiber is multi-clad and S-type. Our evaluation consists of two parts: bench-top measurements using a 405~nm picosecond laser to determine the decay time and attenuation length, and a beam test using a 3~GeV/$c$ electron beam to evaluate the light yield and time resolution in a realistic scintillator setup.
\begin{table}
    \caption{Characteristics of the BCF-XL series~\cite{1} and Y-11~\cite{y11_site} from the specification sheets.}
    \makebox[\textwidth][c]{
    \newcolumntype{C}[1]{>{\centering\arraybackslash}p{#1}}
    \begin{tabular}{l||C{2.6cm}|C{2.6cm}|C{2.6cm}|C{1.5cm}}\hline
        & BCF-92XL & BCF-9929AXL & BCF-9995XL & Y-11 \\ \hline \hline
        Absorption Peak (nm)  & 410 & 410 & 345 & 430\\
        Emission Peak (nm)   & 492 & 492 & 450& 476 \\
        Decay Time (ns)  & 2.7 & 2.7 & 2.7 & 6.9\\
        Attenuation Length (m)  & $>$4 & $>$4 & $>$4 & $>$3.5\\ \hline
    \end{tabular}
    }
    \label{char_fibers}
\end{table}

%% file: src/02_laser.tex
\begin{figure}[htb]
  \centering
  \begin{minipage}[t]{1\textwidth}
    \centering
    \vspace{0cm} 
    \includegraphics[width=0.75\linewidth]{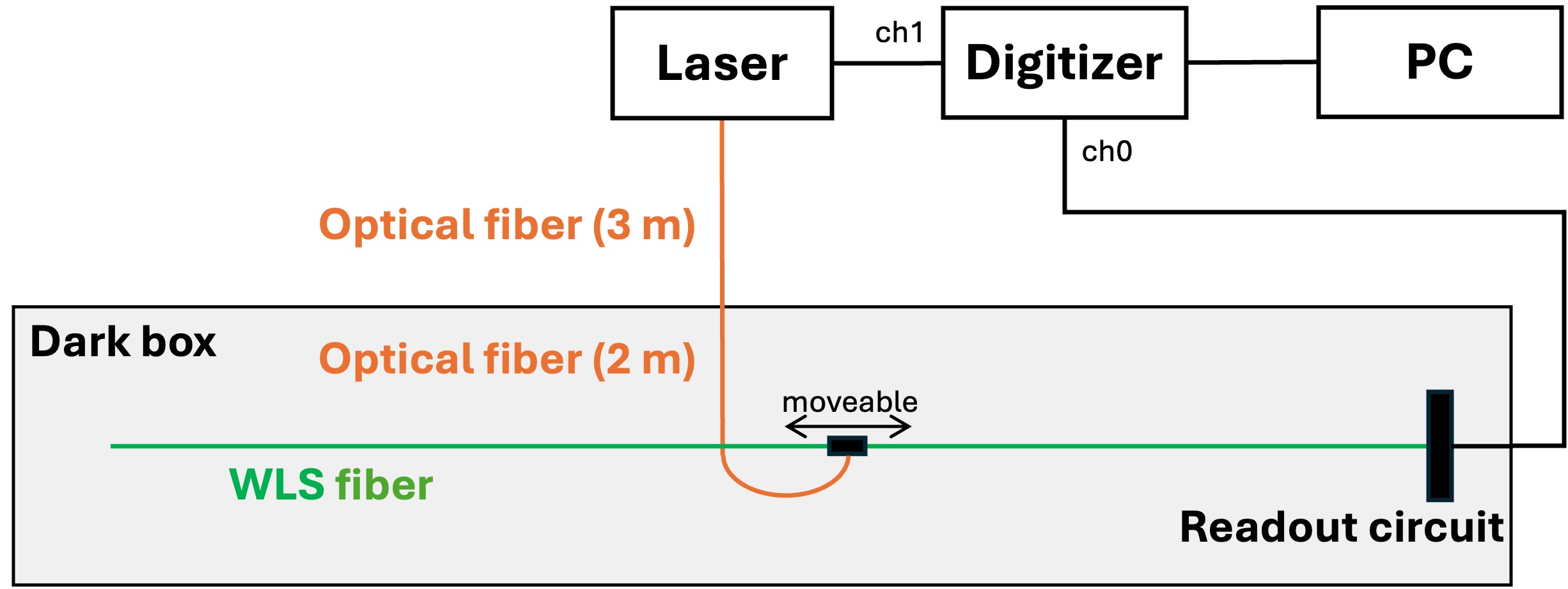}
  \end{minipage}
  \par\vspace{0.2cm}
  \begin{minipage}[t]{1\textwidth}
    \centering
    \vspace{0cm} 
    \includegraphics[width=0.8\linewidth]{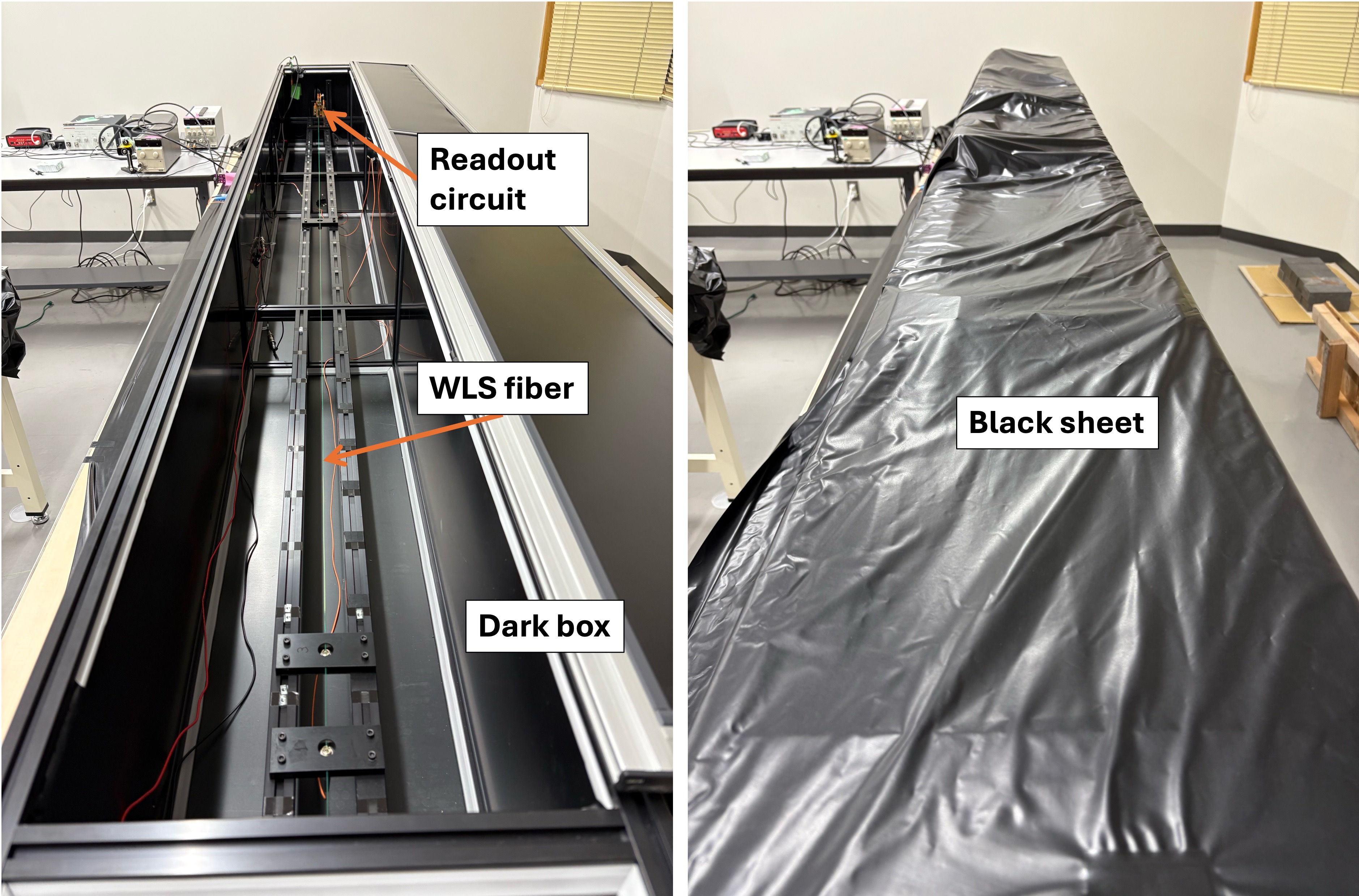}
  \end{minipage}
  \caption{Schematic diagram (top) and photographs (bottom) of the laser-based measurement system. The main components include a laser diode, a digitizer, and a readout circuit on which an MPPC is mounted. After the dark box is closed, it is covered with a black sheet to ensure complete light shielding.}
  \label{setup laser}
\end{figure}

\section{Measurement of decay time and attenuation length using a laser}
\subsection{Setup}
\begin{figure}[htb]
  \centering
  \begin{minipage}[b]{0.59\textwidth}
    \centering
    \vspace{0cm} 
    \includegraphics[width=1\linewidth]{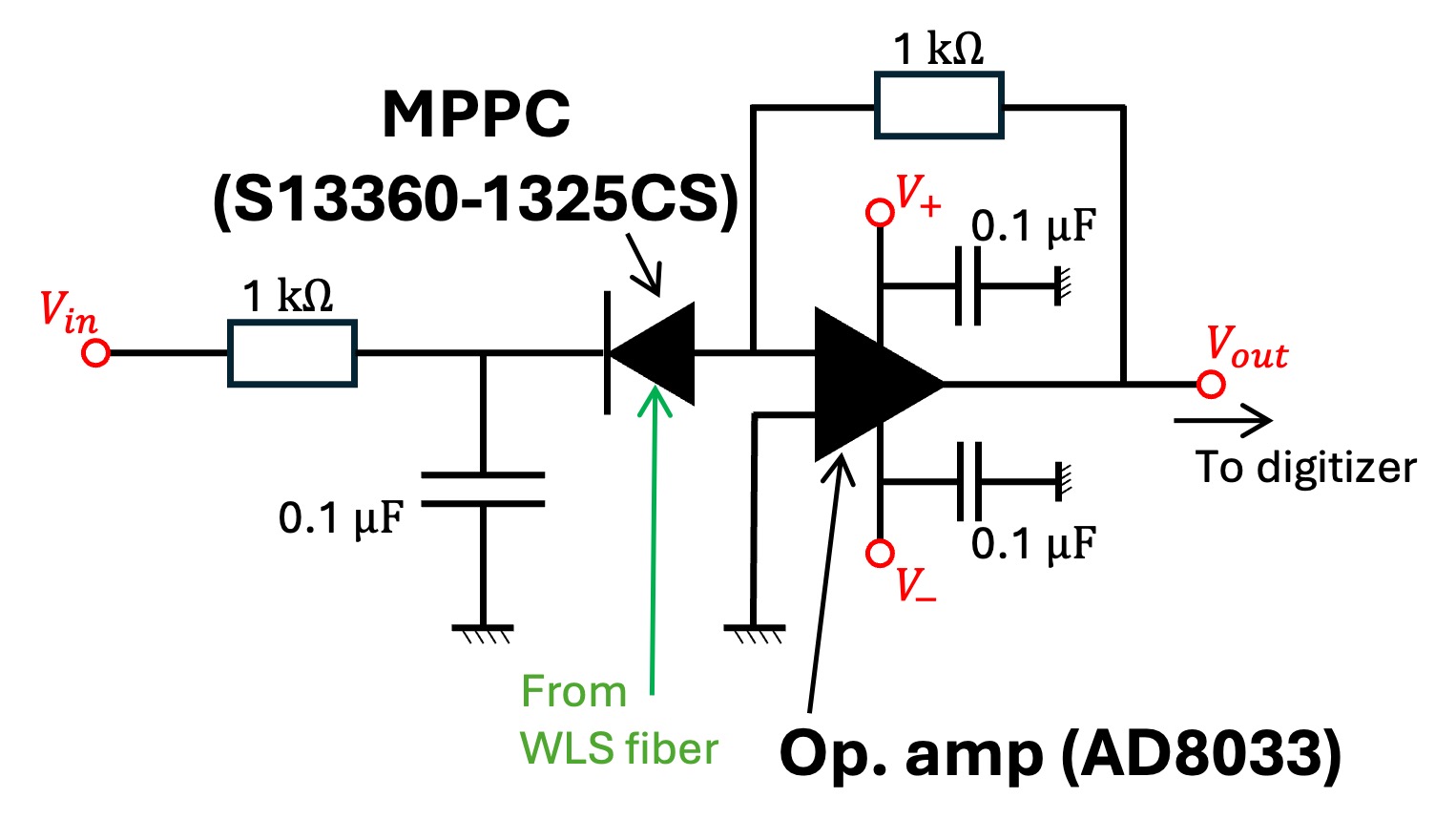}
  \end{minipage}
  \hfill 
  \begin{minipage}[b]{0.40\textwidth}
    \centering
    \vspace{0cm} 
    \includegraphics[width=0.95\linewidth]{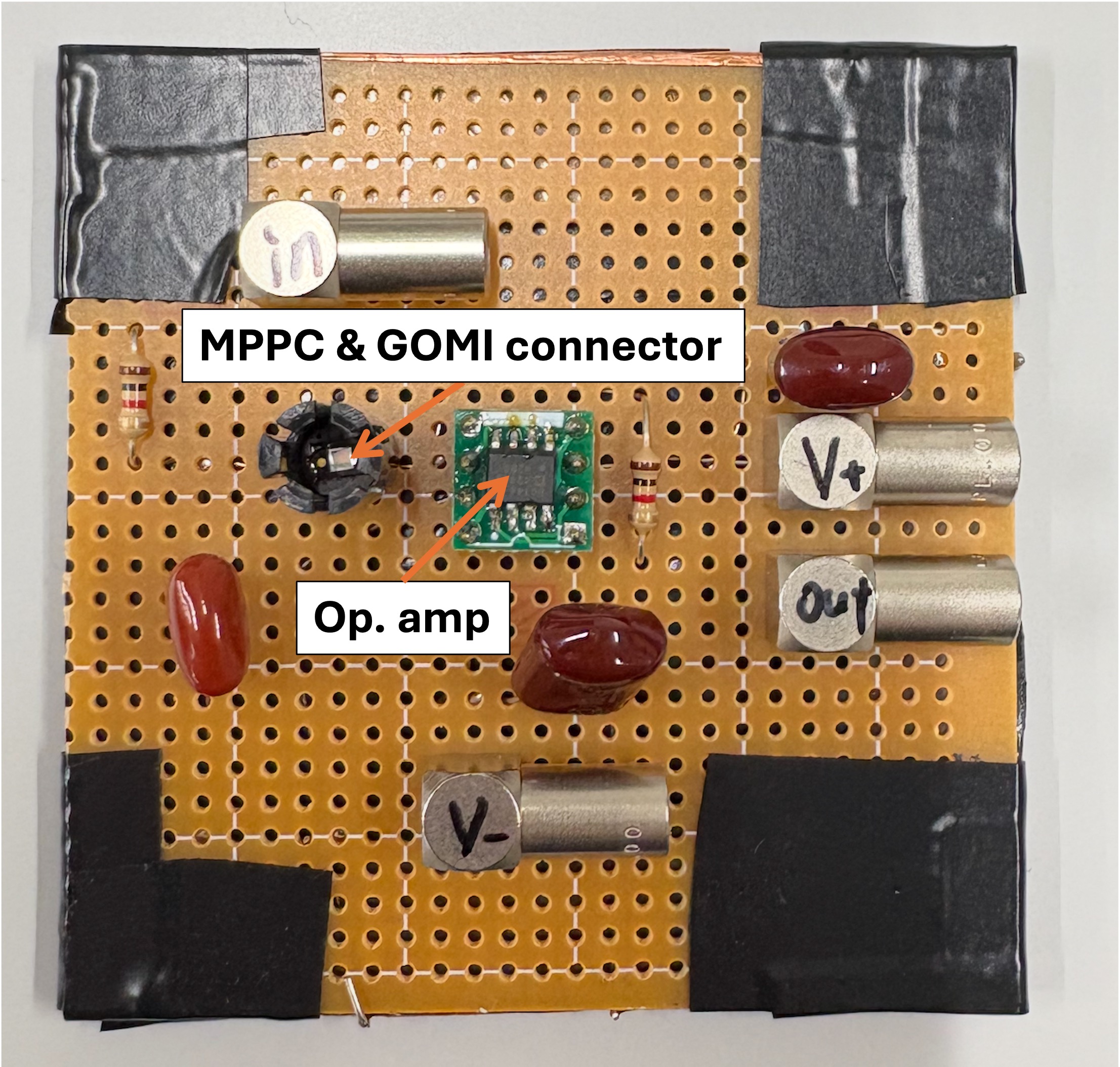}
  \end{minipage}
  \caption{Schematic diagram (left) and photograph (right) of the readout circuit. $V_{in}$ is the bias voltage applied to the MPPC, which is set to approximately 57.5~V.}
  \label{readout}
\end{figure}

\begin{figure}[htb]
  \centering
  \begin{minipage}[b]{0.19\textwidth}
    \centering
    \vspace{0cm} 
    \includegraphics[width=1\linewidth]{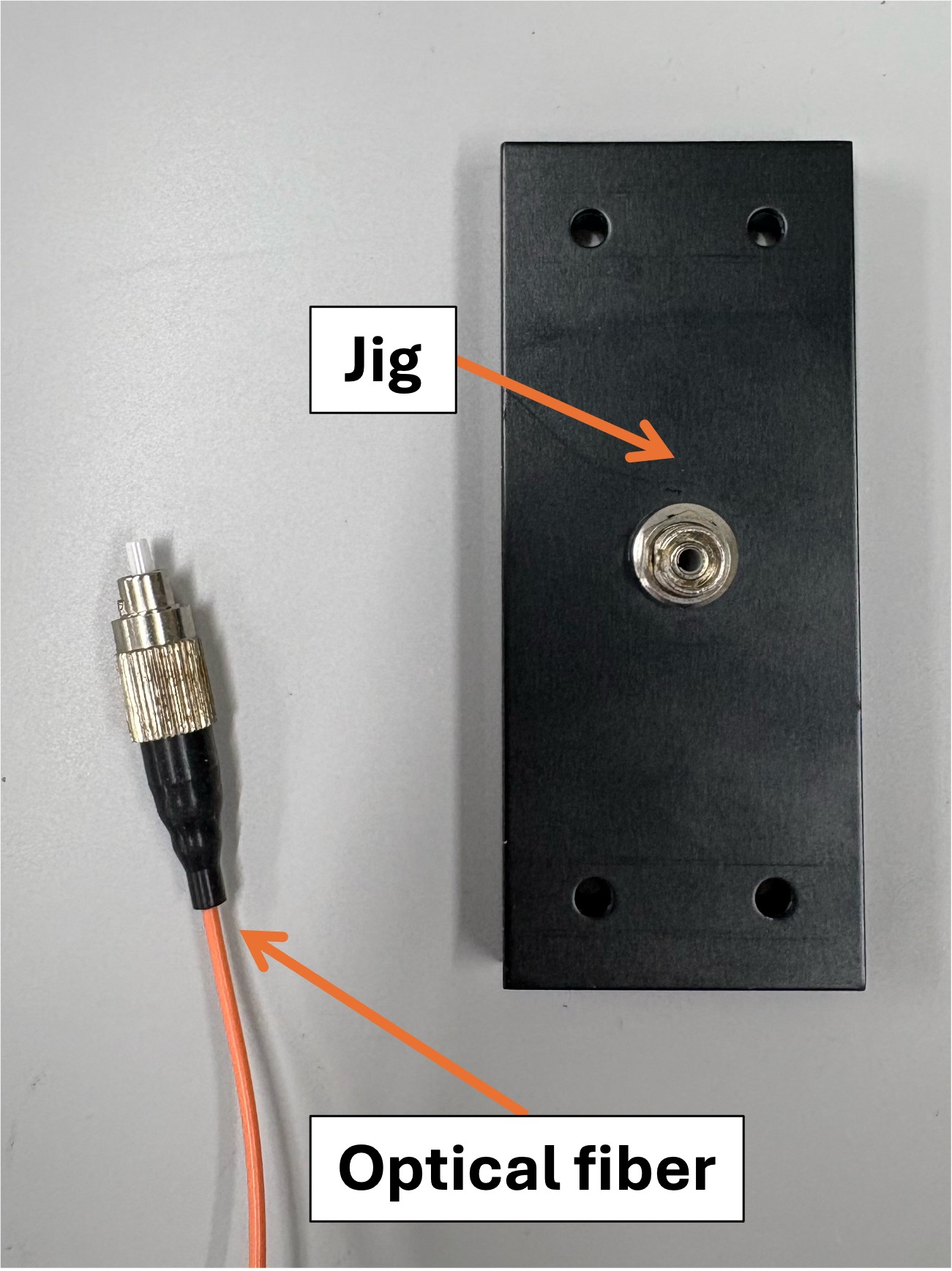}
  \end{minipage}
  \hfill 
  \begin{minipage}[b]{0.19\textwidth}
    \centering
    \vspace{0cm} 
    \includegraphics[width=1\linewidth]{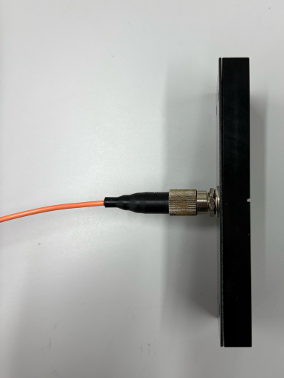}
  \end{minipage}
  \hfill 
  \begin{minipage}[b]{0.19\textwidth}
    \centering
    \vspace{0cm} 
    \includegraphics[width=1\linewidth]{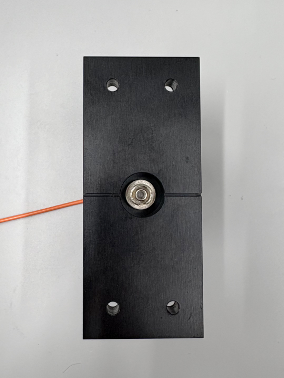}
  \end{minipage}
  \hfill 
  \begin{minipage}[b]{0.19\textwidth}
    \centering
    \vspace{0cm} 
    \includegraphics[width=1\linewidth]{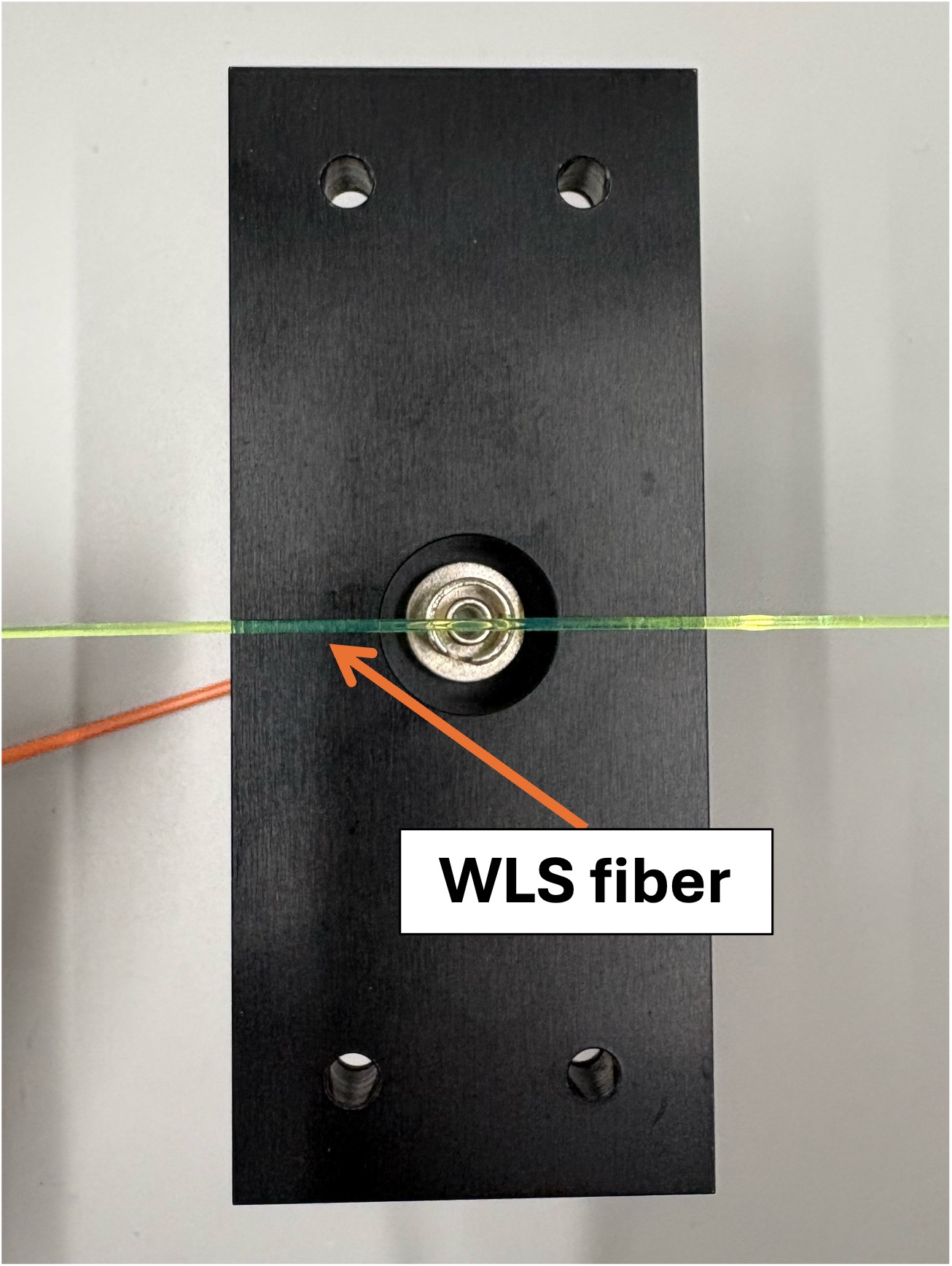}
  \end{minipage}
  \hfill 
  \begin{minipage}[b]{0.19\textwidth}
    \centering
    \vspace{0cm} 
    \includegraphics[width=1\linewidth]{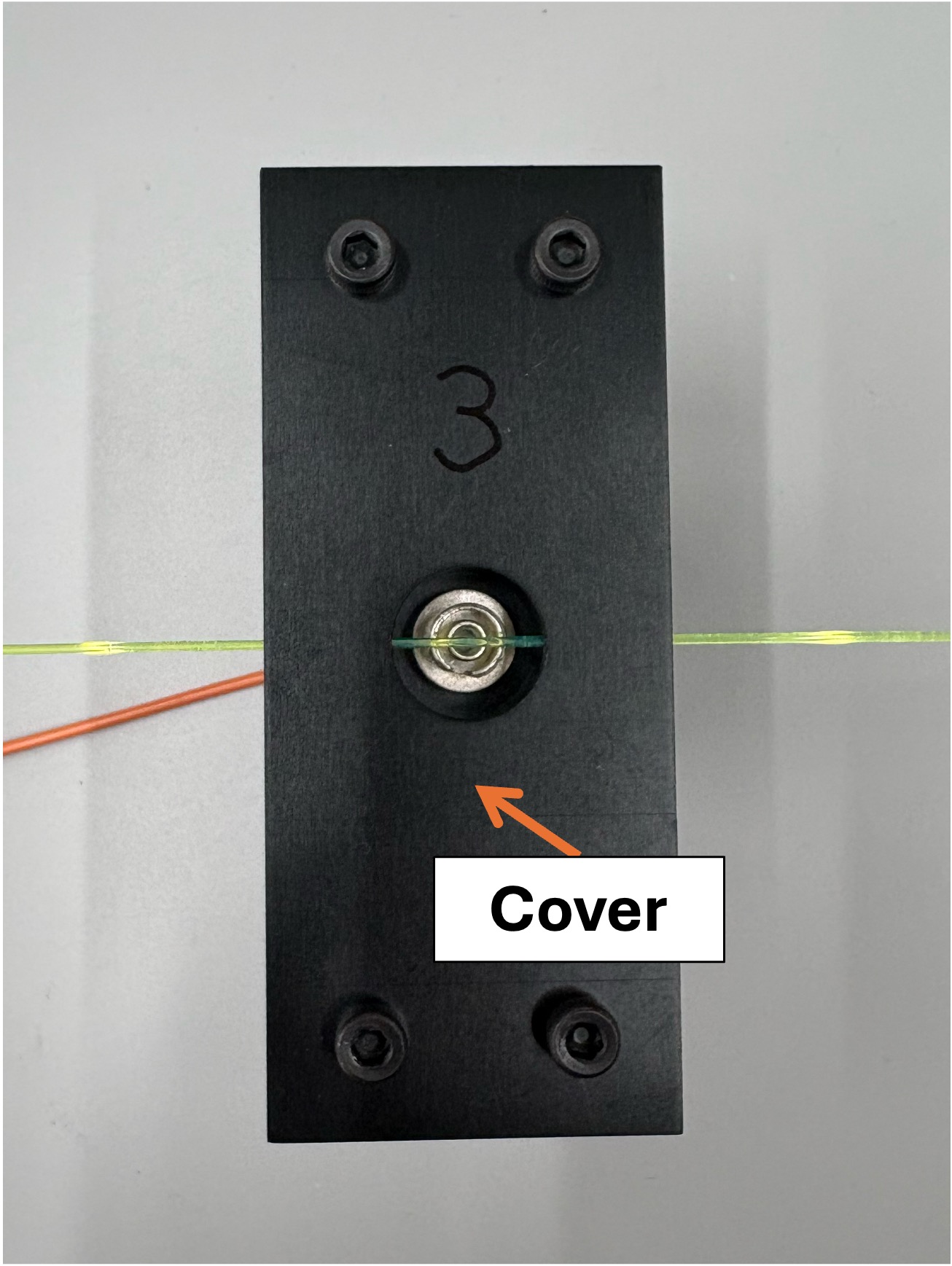}
  \end{minipage}
  \caption{Photographs of the laser injection assembly. The optical fiber is connected to the jig via an FC connector. The WLS fiber is embedded in the groove of the jig and is secured by the cover.}
  \label{injection_position}
\end{figure}

\noindent Figure~\ref{setup laser} shows a schematic diagram of a laser-based measurement system. A WLS fiber is placed inside a dark box. One end of the WLS fiber is coupled to an MPPC (S13360-1325CS, Hamamatsu Photonics), which is mounted on a readout circuit shown in Fig.~\ref{readout}. The opposite end of the fiber is painted black to suppress reflections.

A laser diode (PLP-10, Hamamatsu Photonics) with a wavelength of 405~nm and a pulse width with 60~ps is used as the light source. The light is guided through optical fibers and injected perpendicularly onto the WLS fiber surface. Figure~\ref{injection_position} presents photographs of the laser injection assembly. The optical fiber terminates in an FC connector, allowing it to be connected to a positioning jig. The jig features a groove designed to accommodate the WLS fiber. After placing the WLS fiber in the groove, a cover is placed on top to fix it in position. This injection position is adjustable from 10~cm up to 320~cm from the MPPC. A GOMI connector~\cite{GOMI} is used to couple the WLS fiber to the MPPC. The fiber is glued to the connector plug using optical cement (EJ-500, Eljen Technology~\cite{cement}) and polished to a mirror finish.

The signal from the MPPC is recorded using an 8-channel digitizer (DT5730S, CAEN) with 14-bit resolution and a 500~MS/s sampling rate. A synchronization signal from PLP-10 is connected to another input channel of the digitizer and is used as the trigger for data acquisition.

\subsection{Decay time}
\begin{figure}[htb]
    \centering
    \includegraphics[width=0.95\linewidth]{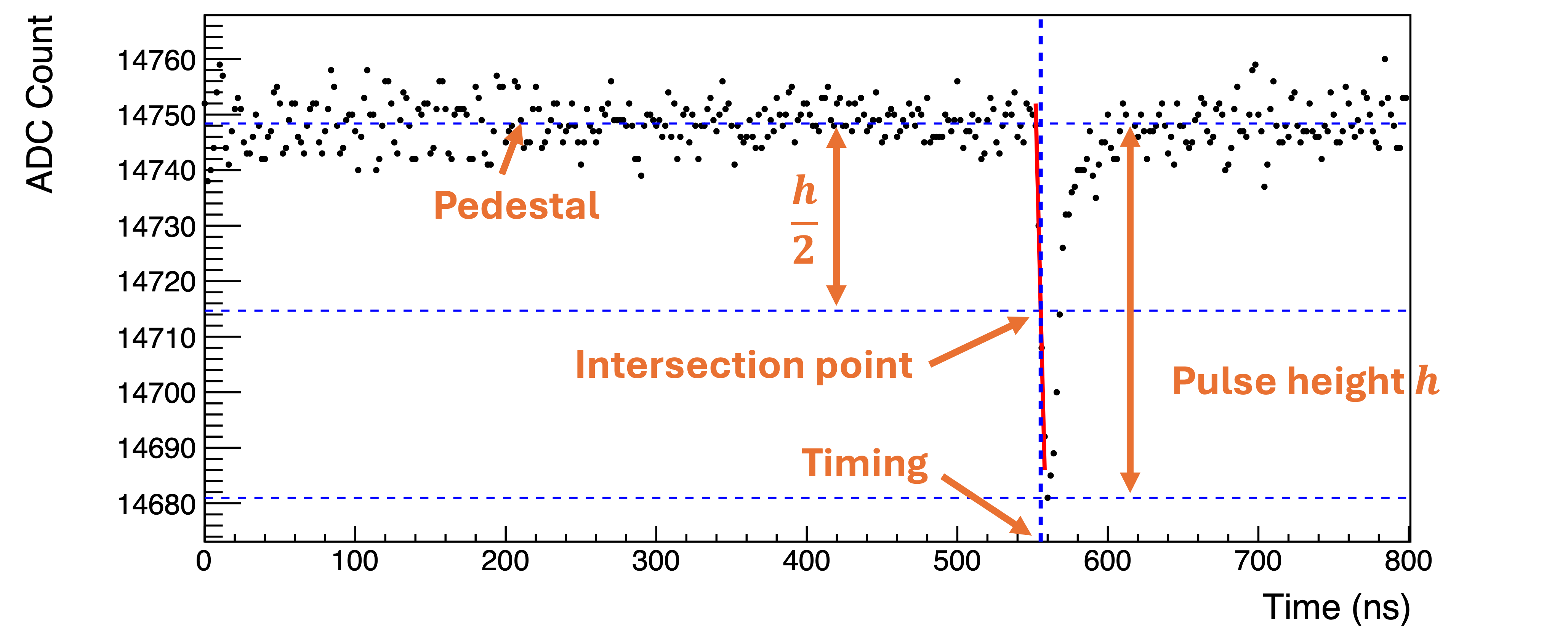}
    \caption{A waveform of a single-photon event. The horizontal axis shows the time (ns), and the vertical axis shows the ADC count of the digitizer. The dashed blue horizontal lines show the pedestal, the half-maximum level, and the pulse height. The red line is the interpolation line used to determine the waveform timing.}
    \label{waveform from the MPPC}
\end{figure}
\begin{figure}[htb]
    \centering
    \includegraphics[width=0.7\linewidth]{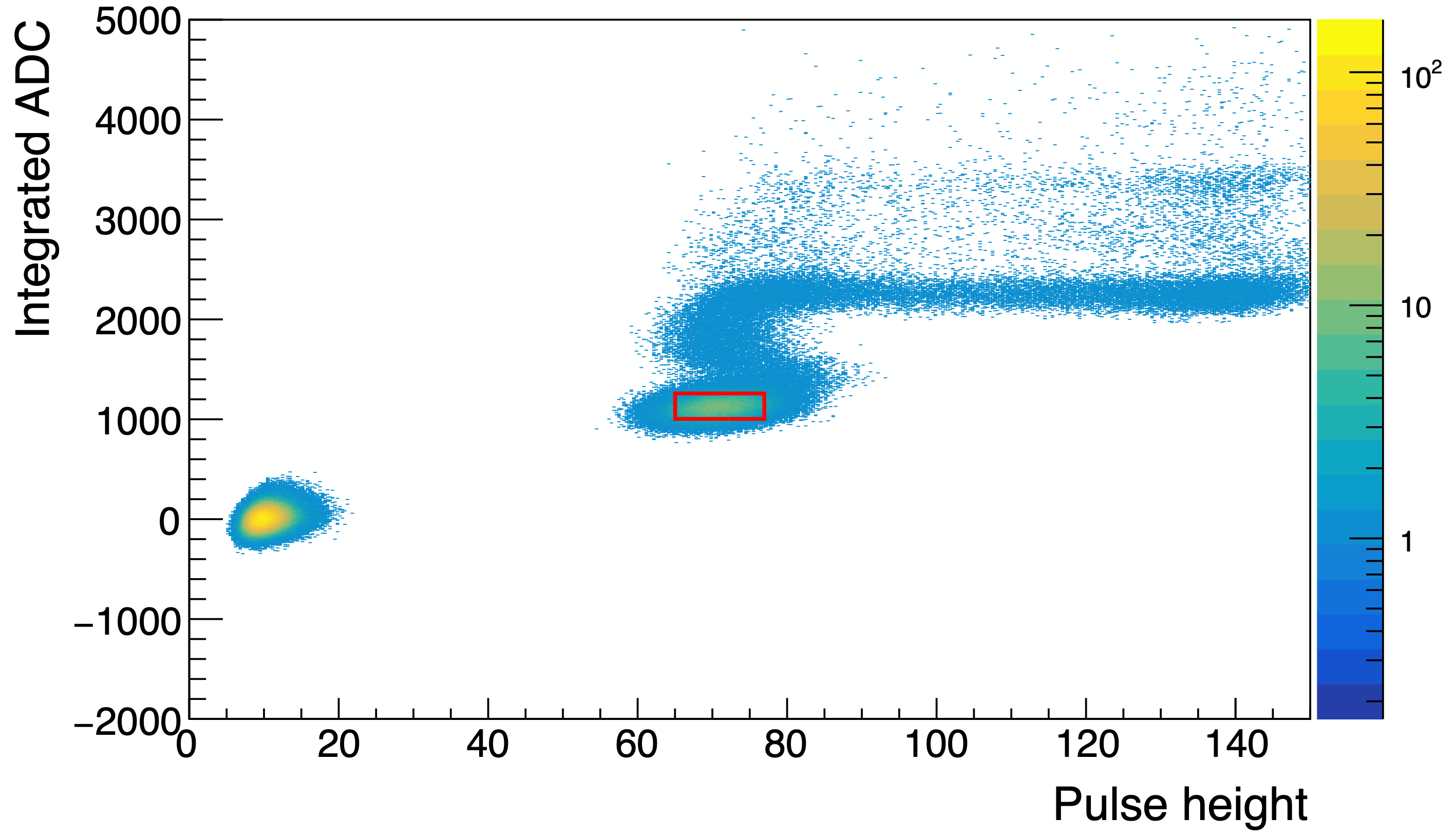}
    \caption{A 2D histogram of pulse height versus integrated ADC. The peak in the lower-left corner corresponds to the 0~p.e. events, while the peak around a pulse height of approximately 70 ADC and an integrated ADC of about 1000 corresponds to the 1~p.e. events. Distributions originating from afterpulses and optical crosstalk can be seen in the upper-right area. The events within the red-boxed region are used for the decay time analysis.}
    \label{eventselection_decaytime}
\end{figure}

\label{decaytime_method}
\noindent For the decay time measurement, we use one sample of each BCF-XL type and seven Y-11 fibers. These seven Y-11 fibers are used to estimate sample-to-sample variations. The length of each fiber is 3~m, and the laser light is injected at a position 10~cm from the MPPC. The laser repetition rate is set to 2~kHz, and the acquisition time for each measurement is 10~min. For the BCF-9995XL, the repetition rate is increased to 10~kHz to accumulate sufficient statistics, as the laser wavelength does not match the fiber's absorption peak.

Each trigger records a waveform of 400 data points, corresponding to an 800~ns time window. Figure~\ref{waveform from the MPPC} shows an example waveform. The pedestal is calculated by averaging the ADC counts within the first 200 points (400~ns). The pulse height is defined as the difference between the pedestal value and the minimum ADC value of the waveform.

To suppress the effect of noise, events are rejected if the difference between the maximum and minimum ADC values within the pedestal calculation window exceeds 30 ADC counts. Single-photon events are selected by excluding contributions from optical crosstalk and afterpulses based on the pulse height and integrated ADC value. The integrated ADC is calculated by summing the pedestal-subtracted ADC counts within a time window from $-50$~ns to $+100$~ns relative to the pulse peak. Figure~\ref{eventselection_decaytime} shows the 2D distribution of the pulse height and integrated ADC value. Gaussian fits to the single-photon peaks in the 1D projections of the pulse height and integrated ADC distributions are used to determine the mean $\mu$ and standard deviation $\sigma$ for each variable. Events within $\pm1.5\sigma$ of $\mu$ for both variables are selected, as indicated by the red box in Fig.~\ref{eventselection_decaytime}. After these selections, the final event sample is reduced to approximately one-tenth or less of the total recorded events.

We determine the timing of the waveform using linear interpolation on its falling edge (shown as a red line in Fig.~\ref{waveform from the MPPC}). The two data points bracketing the half-maximum are used for the interpolation. The timing is then defined as the intersection of this interpolated line with the half-maximum line. We use the same method to determine the timing of the synchronization signal.

\begin{figure}[htb]
  \centering 

  \begin{subfigure}[b]{0.49\textwidth}
    \centering
    \includegraphics[width=\linewidth]{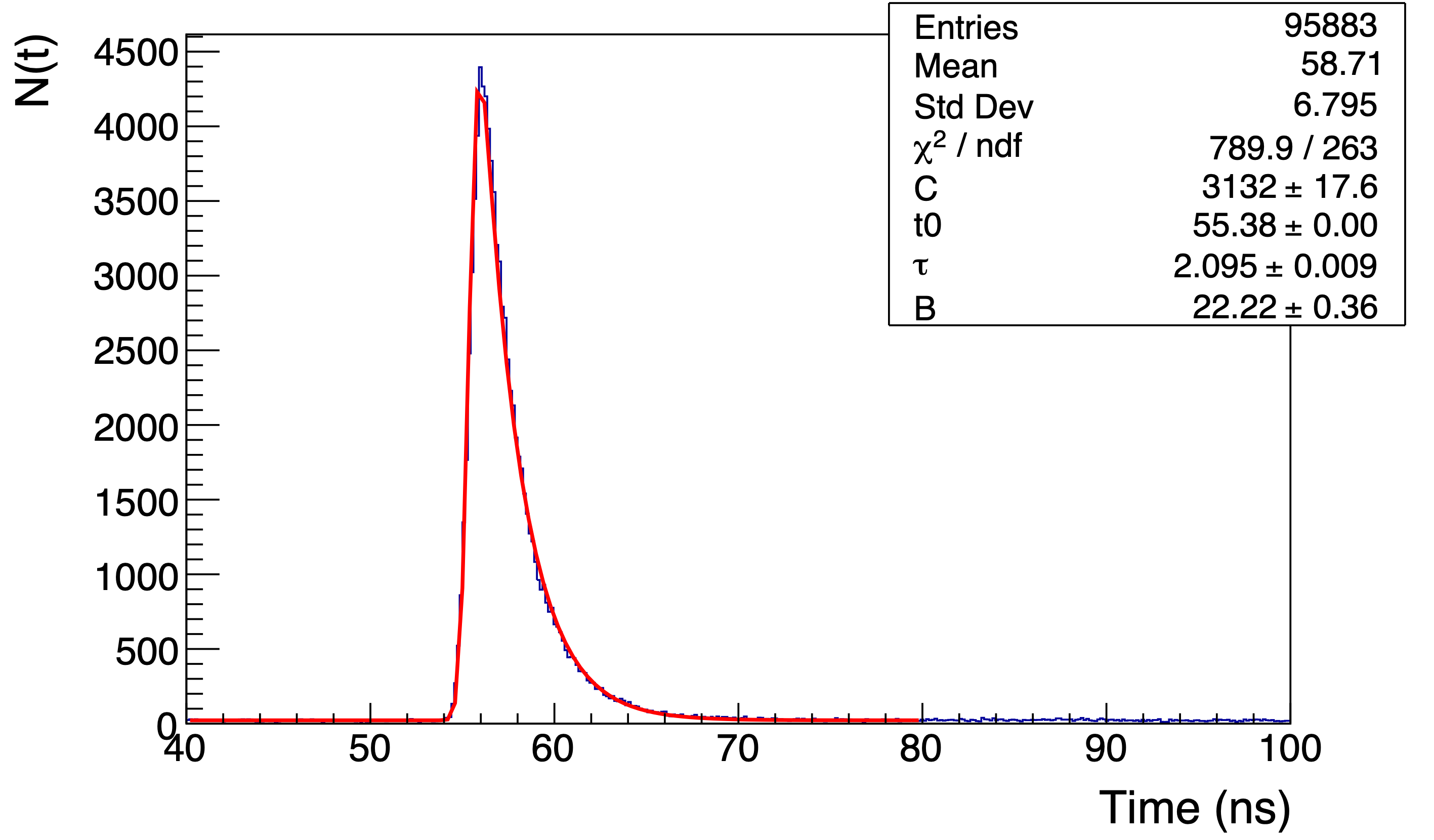}
    \caption{BCF-92XL}
  \end{subfigure}
  \begin{subfigure}[b]{0.49\textwidth}
    \centering
    \includegraphics[width=\linewidth]{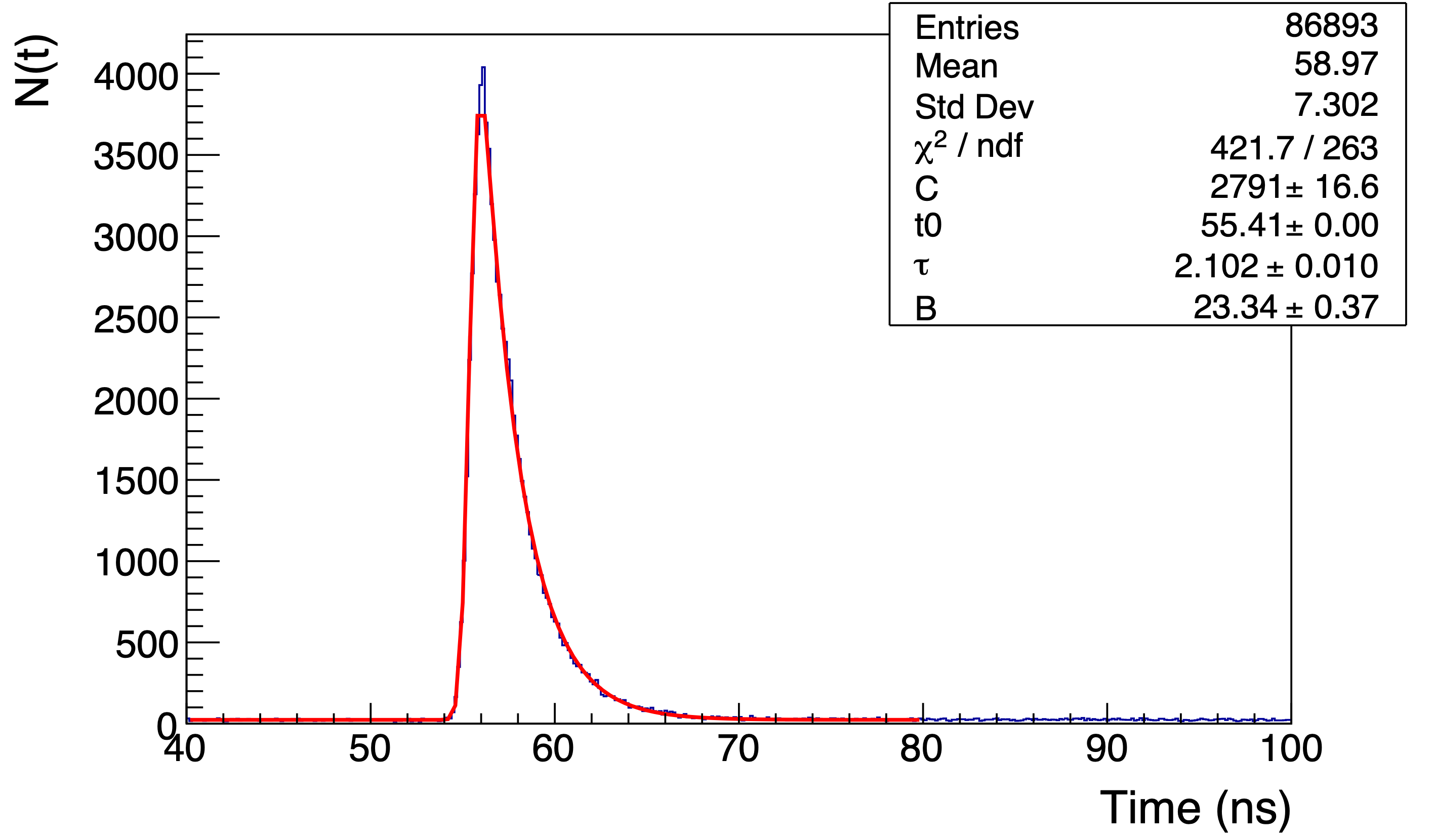}
    \caption{BCF-9929AXL}
  \end{subfigure}
  \\[1em] 
  \begin{subfigure}[b]{0.49\textwidth}
    \centering
    \includegraphics[width=\linewidth]{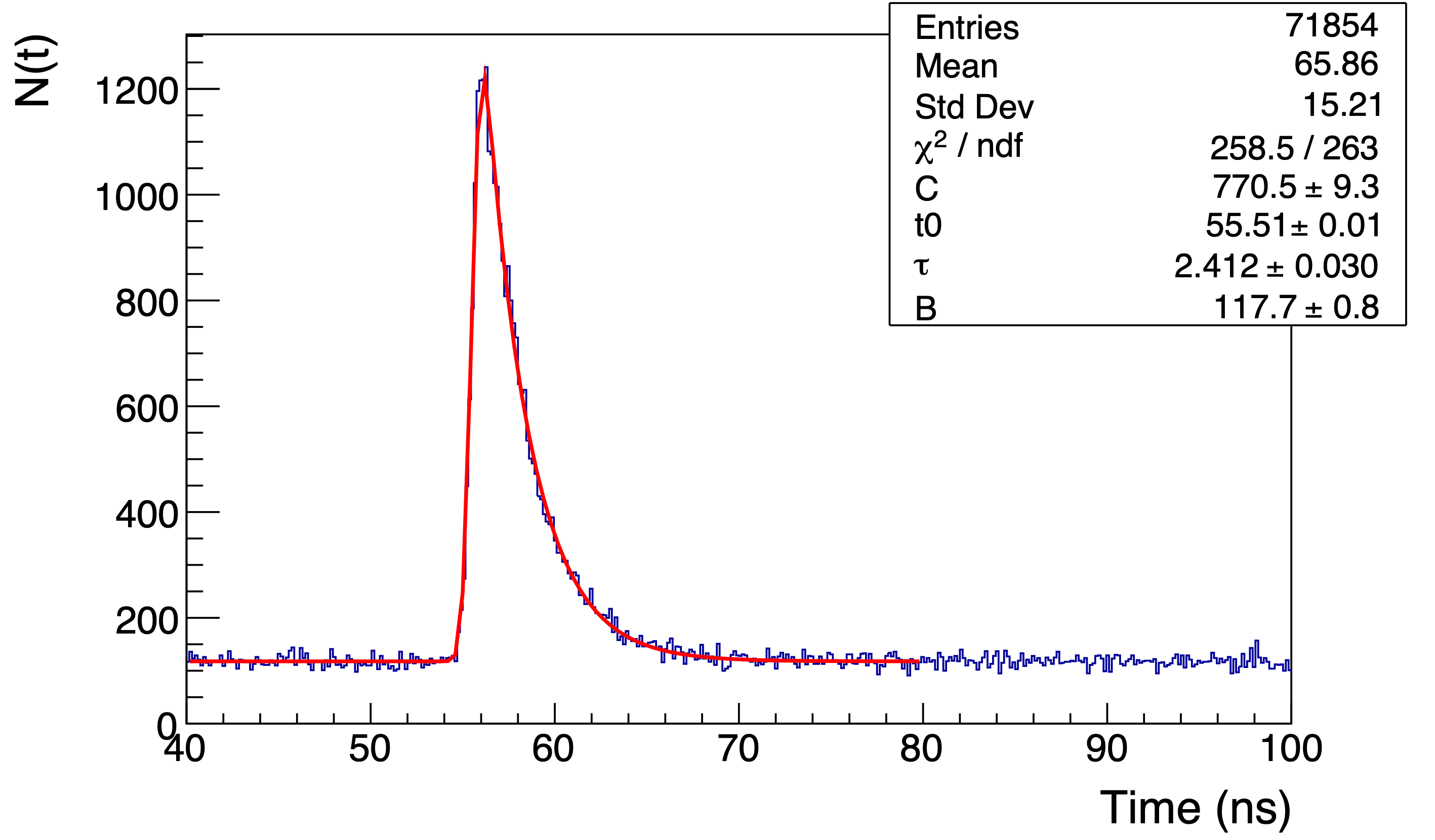}
    \caption{BCF-9995XL}
  \end{subfigure}
  \begin{subfigure}[b]{0.49\textwidth}
    \centering
    \includegraphics[width=\linewidth]{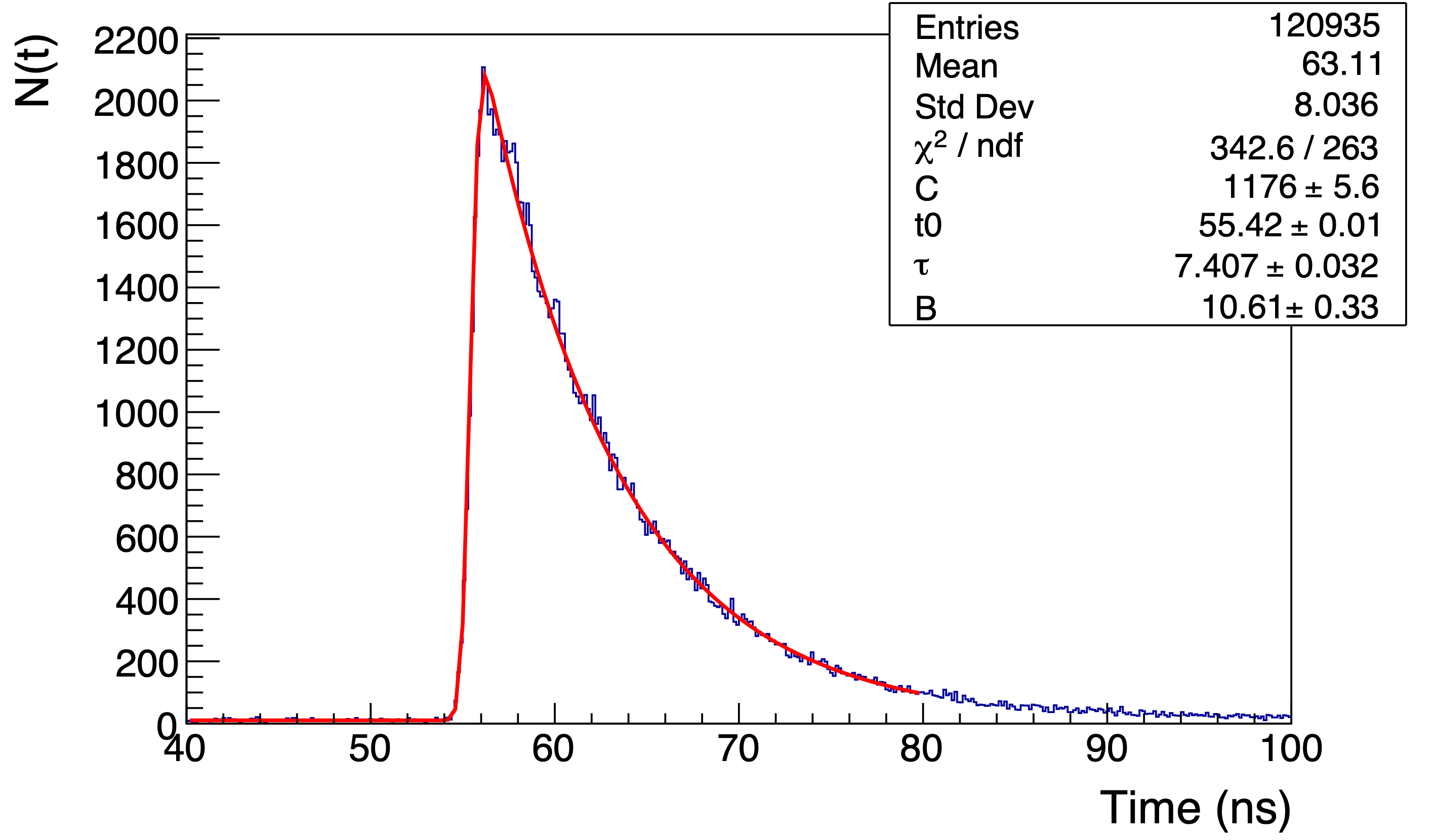} 
    \caption{Y-11}
  \end{subfigure}
  \caption{Distributions of $t$ for each fiber and the corresponding fitting results.}
  \label{decaytime_results}
\end{figure}

The decay time is determined from the distribution of the time difference between the MPPC signal and the synchronization signal, $t$. This distribution is modeled by the function
\begin{equation}
    N(t)=C\left[1+\text{erf}\left(\frac{t-t_0-\frac{\sigma_{\text{sys}}^2}{\tau}}{\sqrt{2}\sigma_{\text{sys}}}\right)\right]\cdot\exp\left(-\frac{t-t_0}{\tau}\right)+B, \label{equation_decaytime}
\end{equation}
where $C$ is the normalization factor, $t_0$ is the time offset of the signal, $\sigma_{\text{sys}}$ is the time resolution of the measurement system, $\tau$ is the decay time, and $B$ is the background term. This equation represents a convolution of an exponential function with decay time $\tau$ and a Gaussian component that represents the system's time resolution. For each fiber, the distribution of $t$ is fitted in the range from 40 to 80~ns with Eq.~(\ref{equation_decaytime}) to extract the decay time $\tau$, where $\sigma$ is fixed to 0.381~ns as determined by direct laser injection into the MPPC. Figure~\ref{decaytime_results} shows the $t$ distributions (blue) and their corresponding fit curves (red) for each fiber.

We estimate the uncertainties for the decay time measurement. The statistical uncertainty is obtained directly from the standard error on the $\tau$ parameter in the fit. For the systematic uncertainties, three sources are considered. The uncertainty in reproducibility arising from sample-to-sample variations (including individual fiber differences, optical fiber coupling, and fiber-to-MPPC coupling) is estimated to be approximately 0.4\% by calculating the standard deviation of the seven Y-11 measurements. BCF-XL fibers are assumed to have a comparable uncertainty. The fit-range dependence is evaluated by varying the fit start point between 40 and 50~ns and the end point between 70 and 100~ns. The uncertainty is defined as half the difference between the maximum and minimum resulting decay times. The impact of the system time resolution $\sigma_{\text{sys}}$ is evaluated by varying its value by $\pm1$ standard deviations of its measured uncertainty ($\pm0.001$~ns). The resulting systematic uncertainty is taken as the maximum difference in the fitted decay time.

\begin{table}[htb]
    \caption{Measured decay times and uncertainties.}
    \makebox[\textwidth][c]{
    \newcolumntype{C}[1]{>{\centering\arraybackslash}p{#1}}
    \begin{tabular}{l|l||c|c|c|C{1.5cm}}\hline
        \multicolumn{2}{c||}{} & BCF-92XL & BCF-9929AXL & BCF-9995XL & Y-11 \\ \hline \hline
        \multicolumn{2}{c||}{Decay Time (ns)} & 2.10 & 2.10 & 2.41 & 7.44 \\ \hline \hline
        \multirow{4}{*}{Uncertainties}  & Fitting  & 0.009 & 0.010 & 0.030 & 0.031 \\ 
        & Reproducibility & 0.009 & 0.009 & 0.010 & 0.031 \\ 
        & Fit Range & 0.006 & 0.006 & 0.003 & 0.040 \\ 
        & System Resolution & 0.001 & 0.001 & 0.001 & 0.001 \\ \hline \hline
        \multicolumn{2}{c||}{Total Uncertainty} & 0.01 & 0.02 & 0.03 & 0.06 \\ \hline
    \end{tabular}
    }
    \label{decaytime_result_table}
\end{table}

The total uncertainty for this measurement is calculated as the quadrature sum of the statistical and systematic uncertainties. The measured decay times, with their statistical and systematic uncertainties, are summarized in Table~\ref{decaytime_result_table}. These results show that the decay times of the BCF-XL series are approximately 30\% of that of the Y-11 fibers.

\subsection{Attenuation length}
\begin{figure}[htb]
  \centering
  \begin{subfigure}[b]{0.3\textwidth}
    \centering
    \vspace{0cm} 
    \includegraphics[width=1\linewidth]{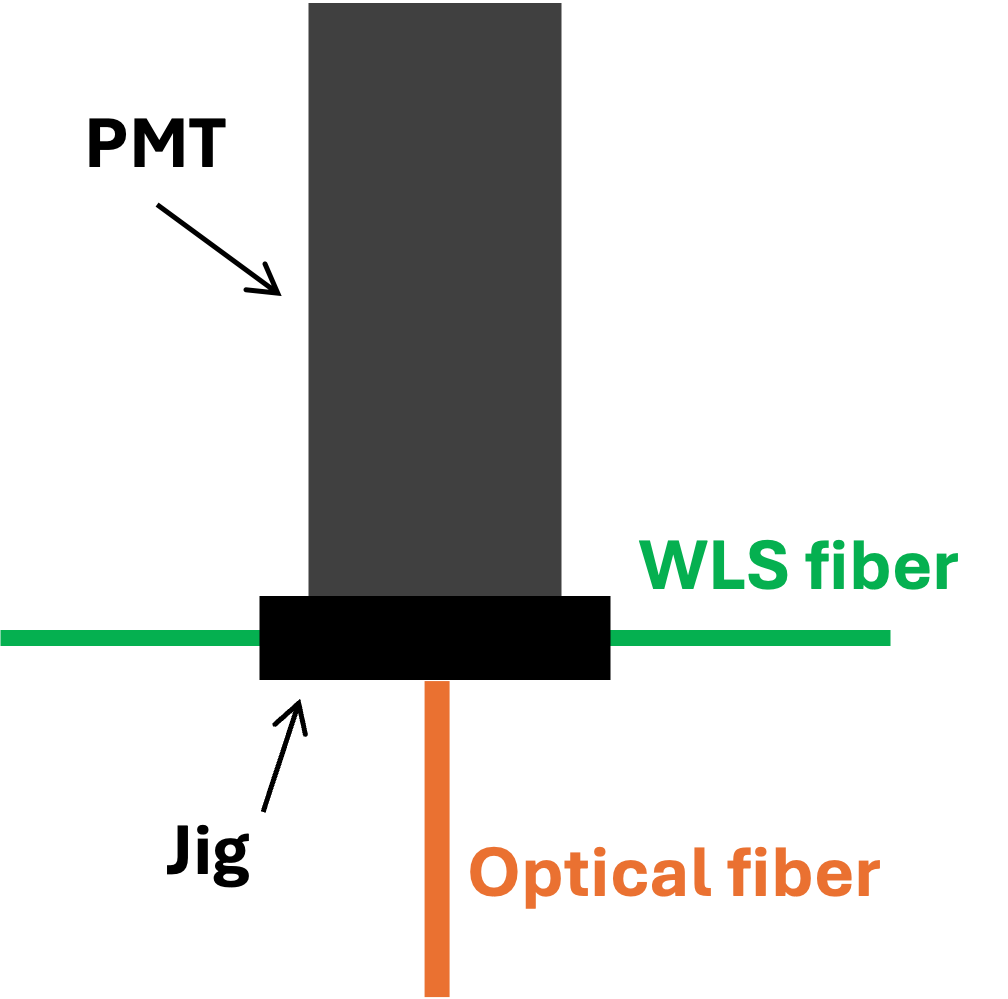}
  \end{subfigure}
  \hfill 
  \begin{subfigure}[b]{0.6\textwidth}
    \centering
    \vspace{0cm} 
    \includegraphics[width=0.9\linewidth]{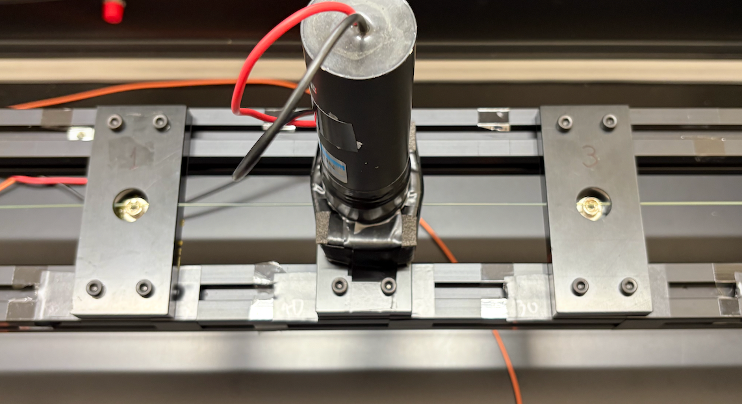}
  \end{subfigure}
  \caption{Schematic diagram (left) and photograph (right) around the laser injection point in the modified setup. A small PMT is mounted on the jig to monitor the laser light stability.}
  \label{modified_setup_atte}
\end{figure}

For the attenuation length measurement, a 3.5~m long fiber of each type is measured. The BCF-9995XL fiber is excluded from this measurement as its light yield is insufficient. To monitor the laser light stability, a PMT (H7416, Hamamatsu Photonics) is attached onto the laser injection jig as shown in Fig.~\ref{modified_setup_atte}. Its signal is also recorded by the digitizer.

The light yield is measured at 27 laser injection positions: in 5~cm steps from 10~cm to 70~cm, in 10~cm steps from 70~cm to 100~cm, and in 20~cm steps from 100~cm up to 320~cm. The laser repetition rate is 10~kHz and the acquisition time is 30~s at each position.

The waveforms are integrated within a time window from $-50$~ns to $+100$~ns relative to the peak time. A Gaussian fit is applied to the light yield distribution at each injection position and the mean is taken as the representative light yield. The light yield for the monitor PMT is also determined in the same way. To correct for variations in the laser intensity, the light yield at each position is multiplied by $M_0/M_i$, where $M_0$ and $M_i$ are the light yields of the monitor PMT at the initial and the corresponding positions, respectively.

We adopt the following two-component exponential function to model the light yield $L(x)$ at a distance $x$ from the MPPC:
\begin{equation}
    L(x)=L_0\cdot\left[\alpha\cdot\exp\left(-\frac{x}{A_L}\right)+(1-\alpha)\cdot\exp\left(-\frac{x}{A_S}\right)\right], \label{attenuation_equation}
\end{equation}
where $L_0$ is the light yield at $x=0$, $\alpha$ is the fractional contribution of the long component, and $A_L$ and $A_S$ are the long and short attenuation lengths, respectively. The representative light yields as a function of distance are fitted with Eq.(\ref{attenuation_equation}) to extract the attenuation parameters.

\begin{figure}[htb]
    \centering
    \includegraphics[width=0.6\linewidth]{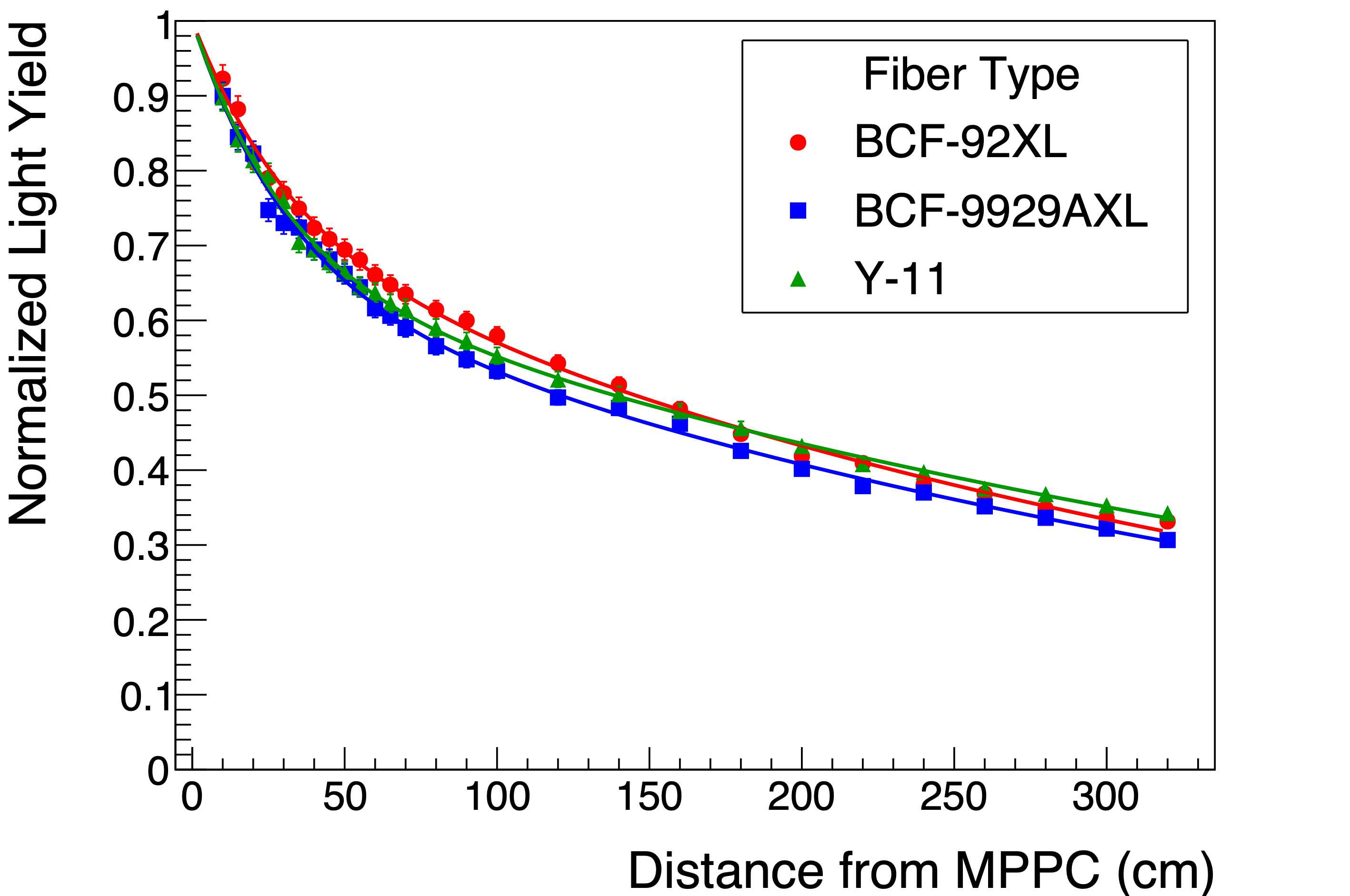}
    \caption{The light yield normalized by $L_0$ as a function of the distance from MPPC for the BCF-92XL (red), BCF-9929AXL (blue), and Y-11 (green) fibers. The solid curves show the results of the two-component exponential fit.}
    \label{attenuation_result}
\end{figure}

\begin{table}[htb]
    \caption{Measured results of attenuation length.}
    \centering
    \newcolumntype{C}[1]{>{\centering\arraybackslash}p{#1}}
    \begin{tabular}{c|C{1.5cm}||C{2.6cm}|C{2.6cm}|C{2.6cm}}\hline
        \multicolumn{2}{c||}{} & BCF-92XL & BCF-9929AXL & Y-11  \\ \hline \hline
        \multirow{3}{*}{Parameters} & $\alpha$ & $0.72\pm0.02$ & $0.66\pm0.02$ & $0.67\pm0.02$ \\ 
         & $A_L$ & $390\pm21$ & $415\pm19$ & $466\pm24$ \\
         & $A_S$ & $32\pm 8$ & $31\pm4$ & $30\pm5$ \\ \hline
    \end{tabular}
    \label{attenuation_table}
\end{table}

The uncorrected light yield shows a systematic fluctuation of approximately 2\%, defined as the standard deviation of ten repeated measurements at the same position. This uncertainty is assumed to be common to all distance points. By correcting the light yield using the monitor PMT, this fluctuation is reduced to about 1\%. Figure~\ref{attenuation_result} shows the light yield normalized by $L_0$ as a function of distance for each fiber. The error bars represent the quadrature sum of this systematic uncertainty and the statistical uncertainty.  

The final parameters obtained from the two-component exponential fit are summarized in Table~\ref{attenuation_table}. From these results, it is found that the attenuation lengths of the BCF-XL series are comparable to that of the Y-11 fiber within the measured distance range.

%% file: src/03_electronbeam.tex
\section{Measurement of light yield and time resolution using an electron beam}

\subsection{Setup}
\begin{figure}[htb]
  \centering
  \begin{minipage}[b]{0.54\textwidth}
    \centering
    \vspace{0cm} 
    \includegraphics[width=1\linewidth]{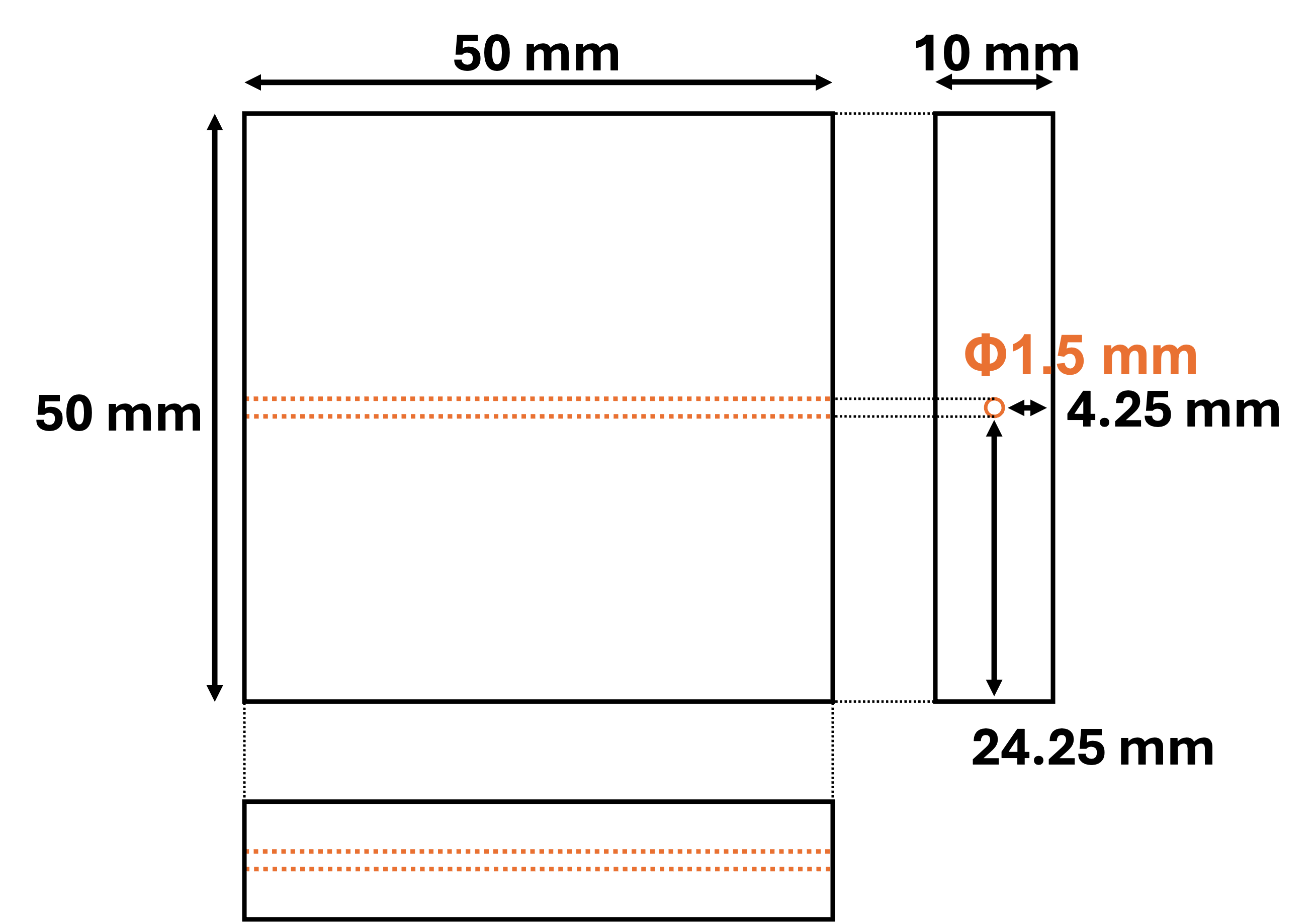}
  \end{minipage}
  \hfill 
  \begin{minipage}[b]{0.42\textwidth}
    \centering
    \vspace{0cm} 
    \includegraphics[width=1\linewidth]{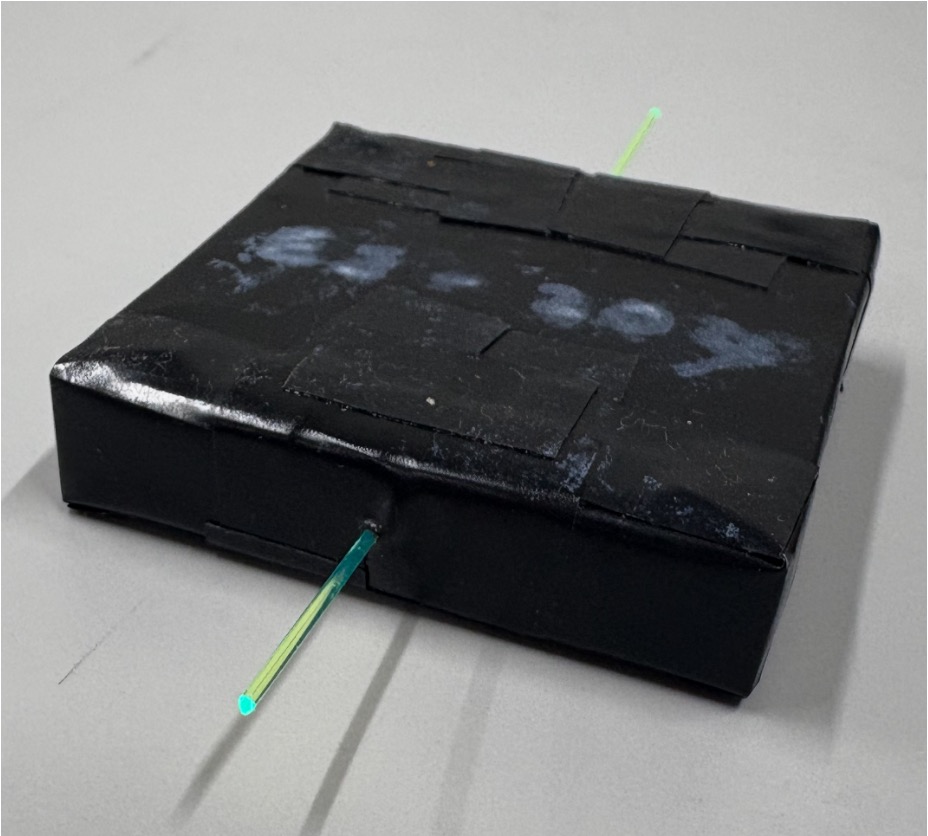}
  \end{minipage}
  \caption{Schematic diagram (left) and photograph (right) of the target scintillator. It is a 50~mm $\times$ 50~mm square with a thickness of 10~mm, featuring a 1.5~mm diameter hole at the center.}
  \label{scinti}
\end{figure}

The light yield and time resolution of the fibers coupled to a plastic scintillator (EJ204, Eljen Technology~\cite{4}) are measured using a 3~GeV/$c$ electron beam at the KEK PF-AR test beamline~\cite{3}. Figure~\ref{scinti} shows the structure of the target scintillator, which is 50~mm $\times$ 50~mm square with a thickness of 10~mm and 1.5~mm diameter hole at the center to accommodate a WLS fiber. The rise time and the decay time of EJ-204 are 0.7~ns and 1.8~ns, respectively \cite{4}. This scintillator is wrapped in aluminum foil as a reflector and covered with black tape for light shielding. Each fiber is inserted into the hole for the measurement.

\begin{figure}[!htb]
  \centering
  \begin{minipage}[t]{1\textwidth}
    \centering
    \vspace{0cm} 
    \includegraphics[width=0.6\linewidth]{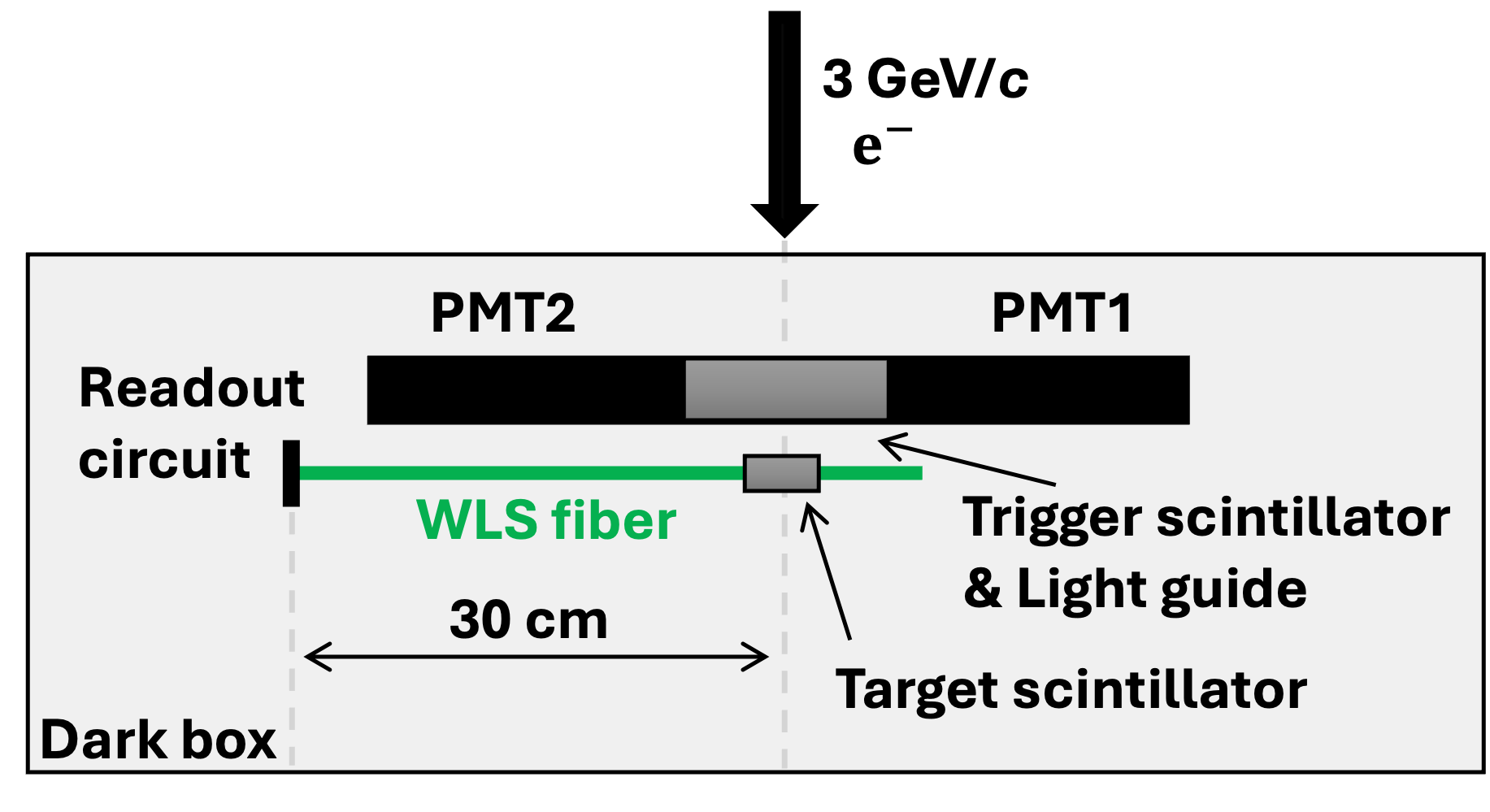}
  \end{minipage}
  \par\vspace{0cm}
  \begin{minipage}[t]{1\textwidth}
    \centering
    \vspace{0cm} 
    \includegraphics[width=0.8\linewidth]{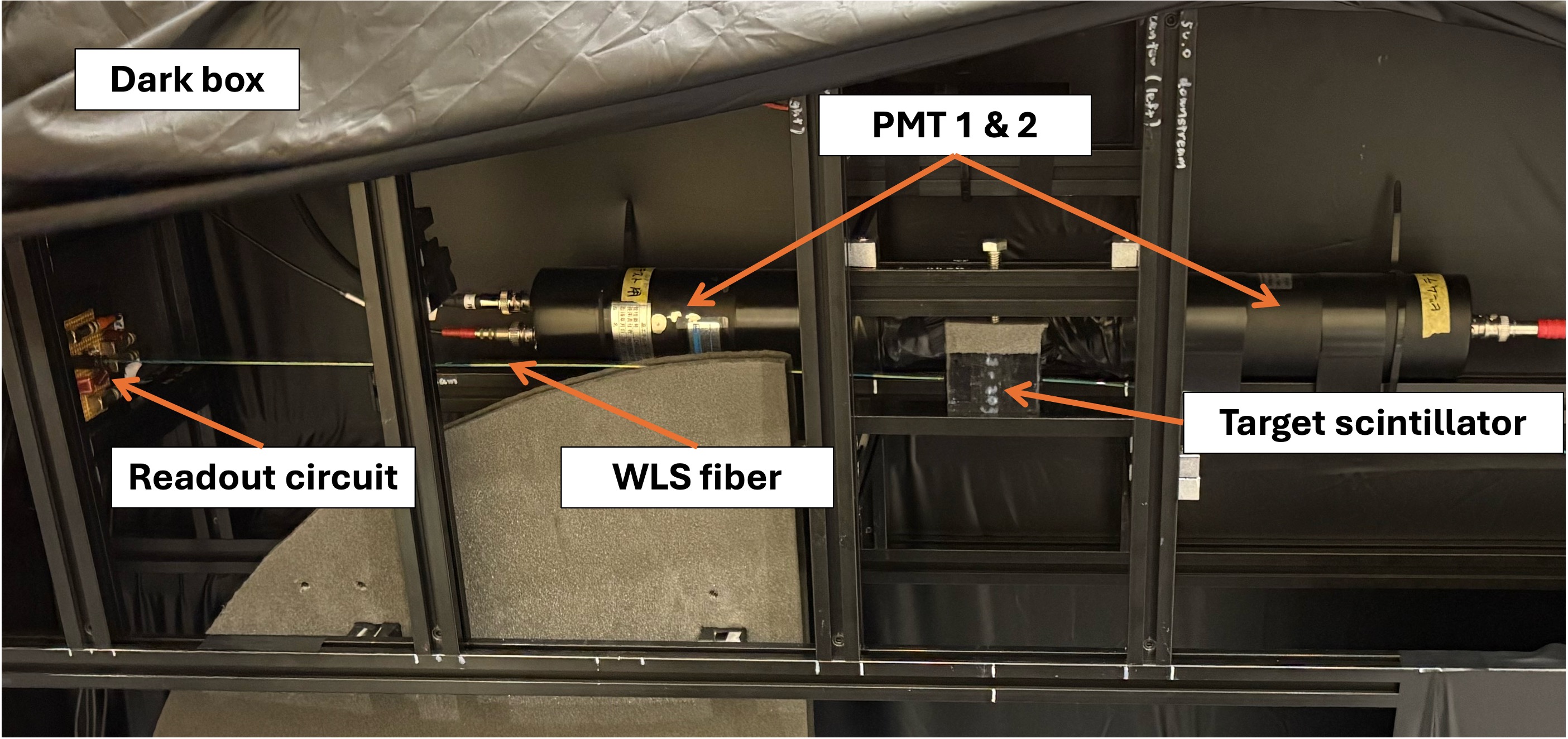}
  \end{minipage}
  \caption{Schematic diagram (top) and photograph (bottom) of the setup using the 3~GeV/$c$ electron beam. The main components include the trigger and target scintillators, two PMTs, the WLS fiber, and the readout circuit.}
  \label{setup_electronbeam}
\end{figure}

\begin{figure}[htb]
    \centering
    \includegraphics[width=1.0\linewidth]{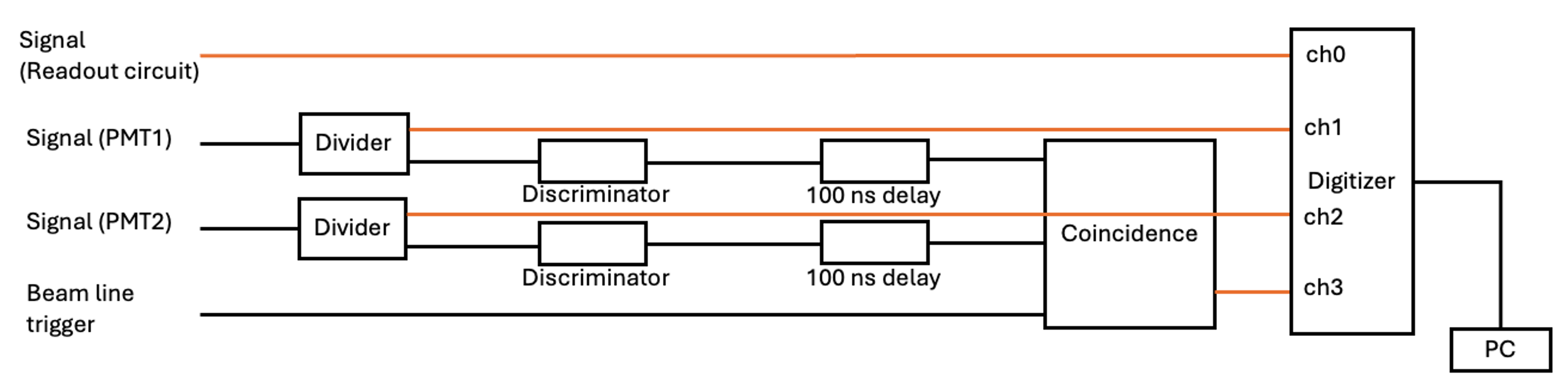}
    \caption{Diagram of the DAQ system. The signals from PMT1 and PMT2 are split by dividers to form a coincidence with the beam line trigger.}
    \label{DAQ}
\end{figure}

Figure~\ref{setup_electronbeam} shows a schematic diagram of the experimental setup.
A trigger scintillator (EJ-204, Eljen Technology~\cite{4}, 40~mm $\times$ 30~mm), read out by PMTs (PMT1 and PMT2) on both sides, is placed upstream of the target scintillator.
The light from the WLS fiber is detected by an MPPC.
The distance between the beam center and the MPPC is 30~cm.
The signals from the MPPC and two PMTs are recorded using the same waveform digitizer (DT5730S, CAEN) as in the laser-based measurements.

Figure~\ref{DAQ} shows a diagram of the data acquisition system.
The signal from the MPPC is fed into channel 0 of the digitizer, while the signals from PMT1 and PMT2 are fed into channels 1 and 2, respectively. 
A coincidence signal is formed from the two PMTs and a trigger signal provided from scintillators placed upstream of the beamline. 
This coincidence signal is fed into channel 3 and used as the trigger for the data acquisition.
The trigger rate is approximately 2.5~kHz.


\subsection{Light yield}
The light yields are measured using seven fibers of each type (BCF-XL series and Y-11). The length of each fiber is 0.4~m. The acquisition time for each measurement is 10~min. A gain calibration using dark pulses is performed immediately before each measurement.

\begin{figure}[htb]
  \centering
  \begin{subfigure}[b]{0.48\textwidth}
    \centering
    \vspace{0cm} 
    \includegraphics[width=1\linewidth]{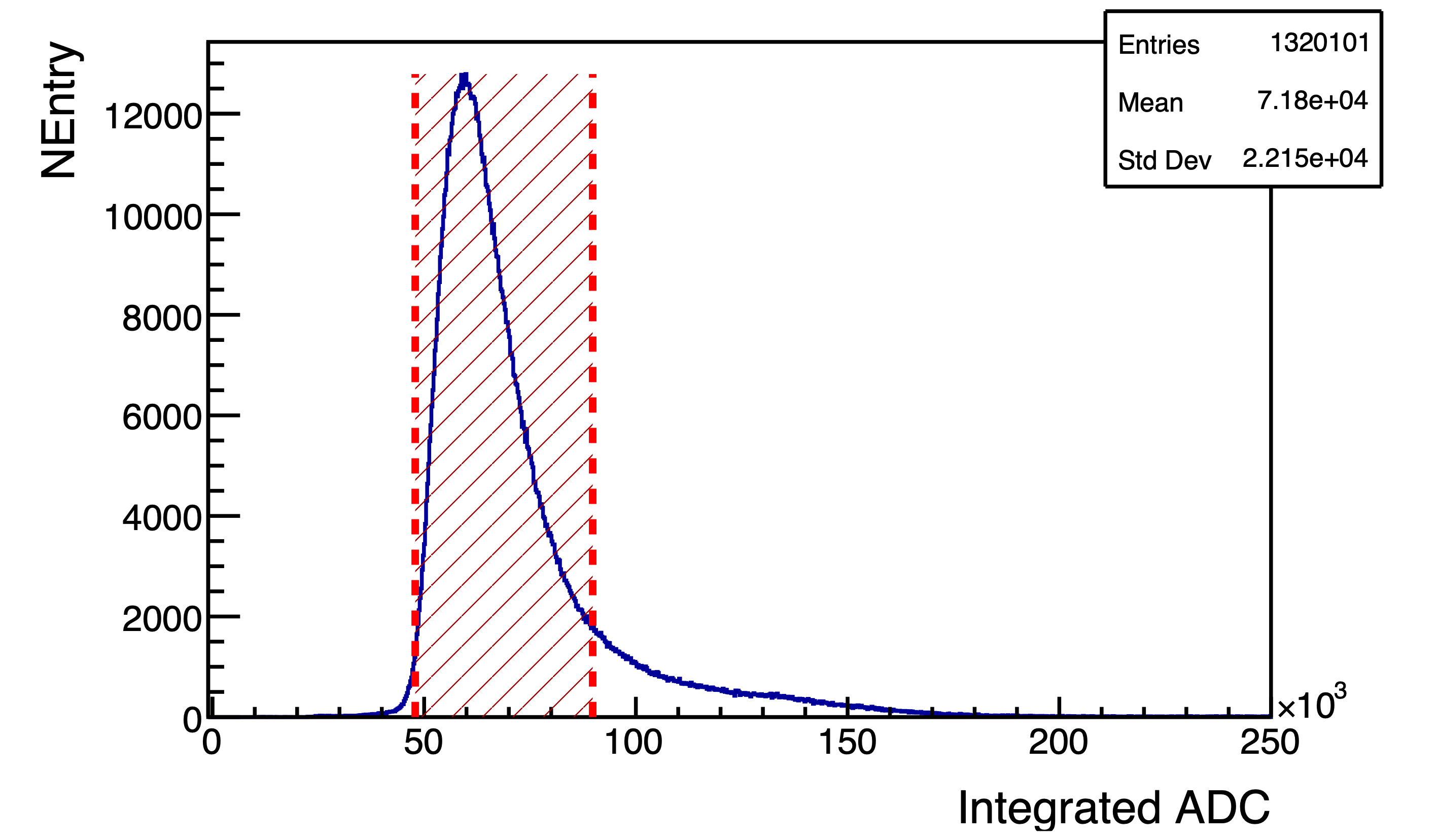}
    \caption{PMT1}
  \end{subfigure}
  \hfill 
  \begin{subfigure}[b]{0.48\textwidth}
    \centering
    \vspace{0cm} 
    \includegraphics[width=1\linewidth]{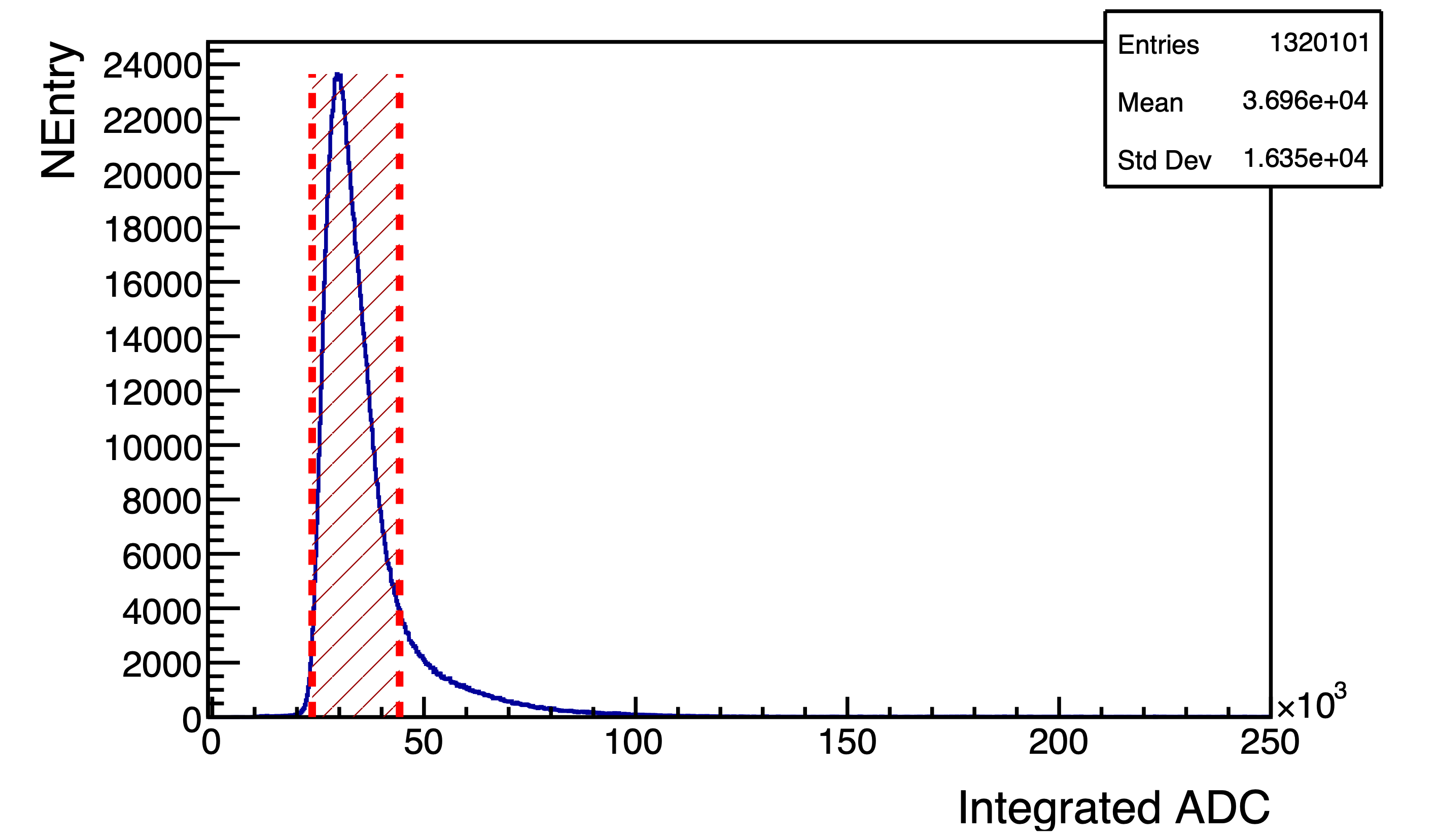}
    \caption{PMT2}
  \end{subfigure}
  \caption{Light yield distributions of the trigger scintillators obtained by PMT1 (left) and PMT2 (right). The horizontal axis shows the integrated ADC, and the vertical axis shows the number of events. The event selection range (0.8 to 1.5 times the peak value) is indicated by the red dashed lines and the hatched region.}
  \label{lightyield_dist_pmt12}
\end{figure}

Figure~\ref{lightyield_dist_pmt12} shows the light yield distributions from the trigger scintillator. 
Events corresponding to a single incident electron are selected by requiring the light yields in both PMTs to be within 0.8 to 1.5 times the peak value, as indicated by the shaded region.

\begin{figure}[htb]
  \centering 

  \begin{subfigure}[b]{0.48\textwidth}
    \centering
    \includegraphics[width=\linewidth]{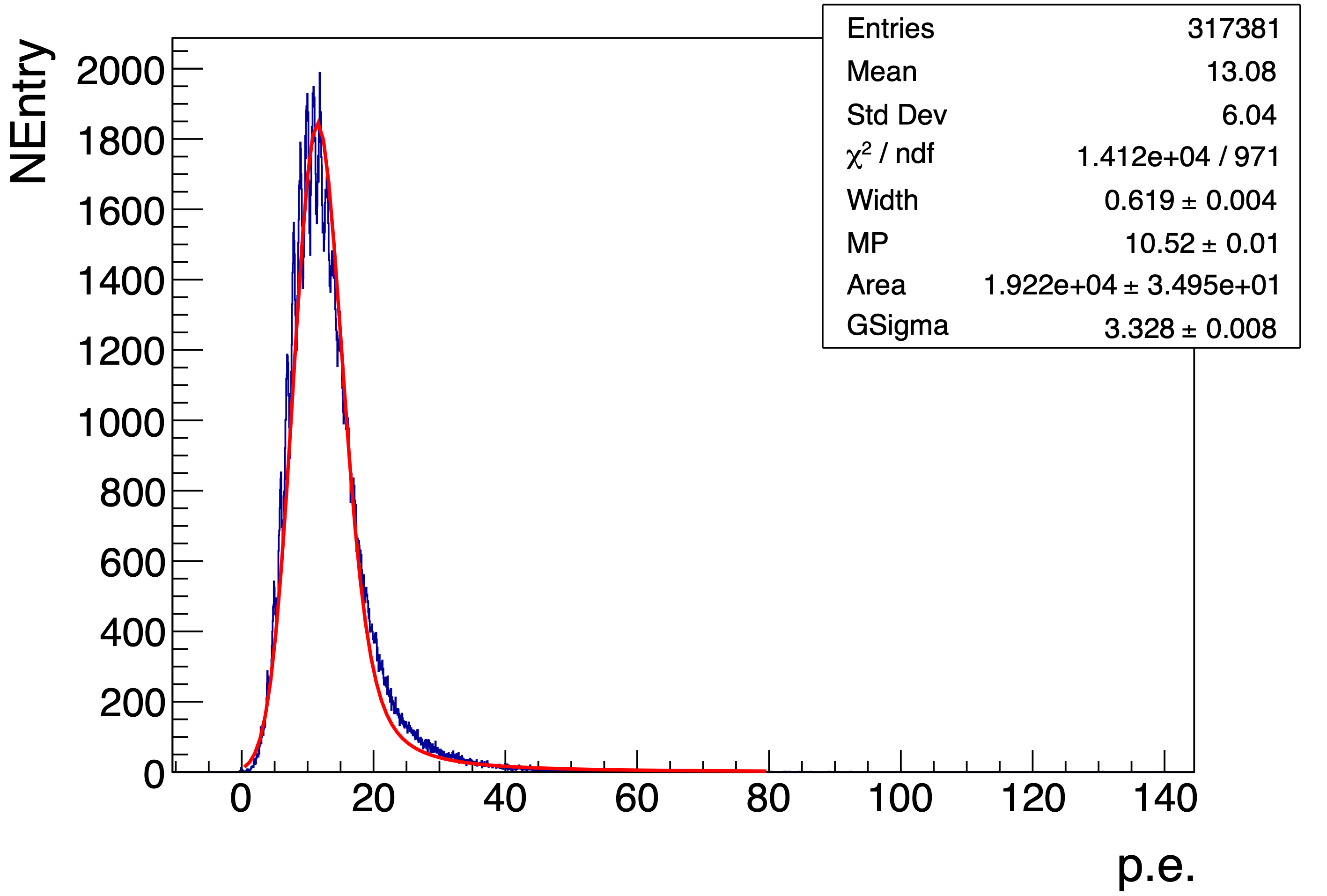}
    \caption{BCF-92XL}
  \end{subfigure}
  \begin{subfigure}[b]{0.48\textwidth}
    \centering
    \includegraphics[width=\linewidth]{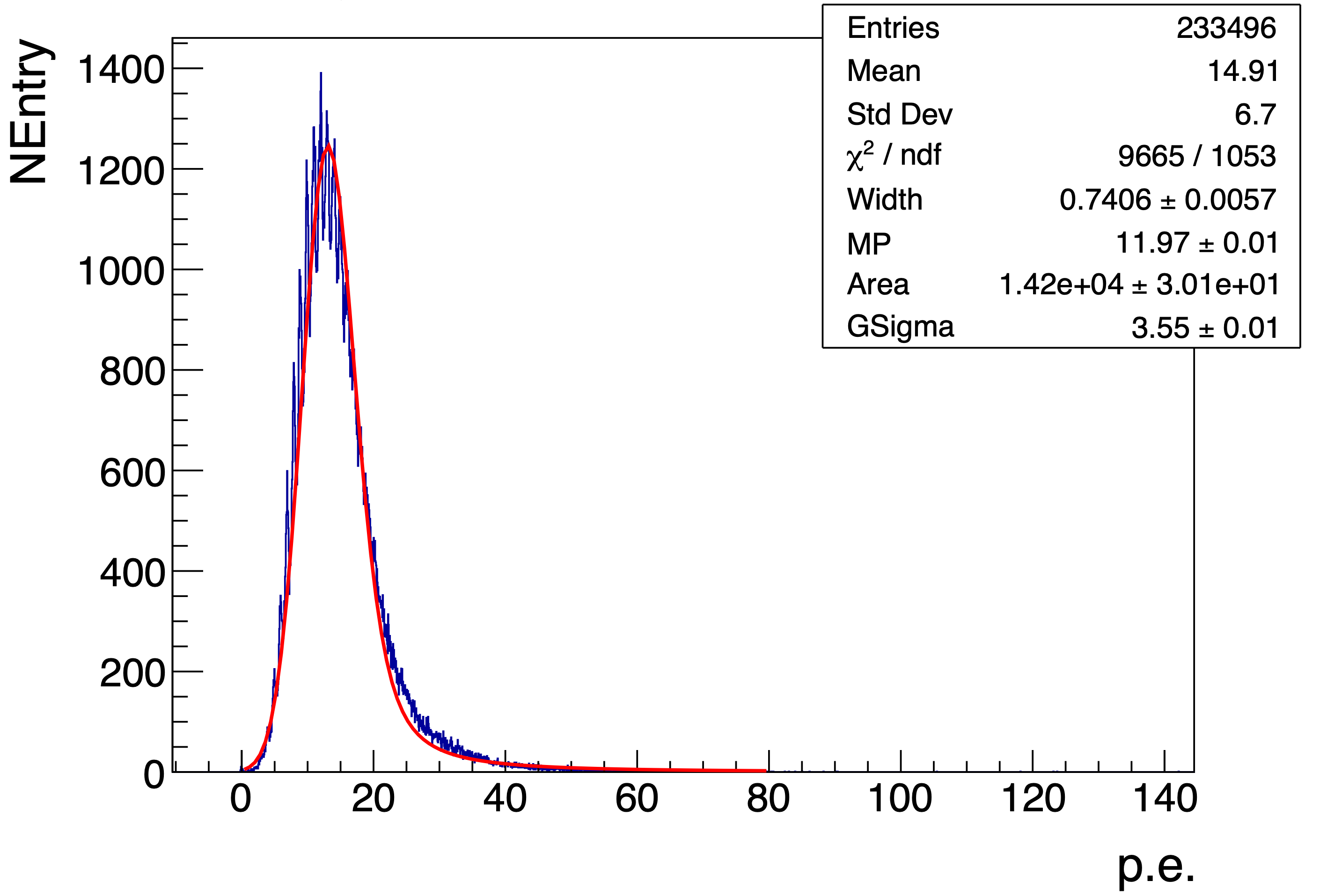}
    \caption{BCF-9929AXL}
  \end{subfigure}
  \\[1em]  
  \begin{subfigure}[b]{0.48\textwidth}
    \centering
    \includegraphics[width=\linewidth]{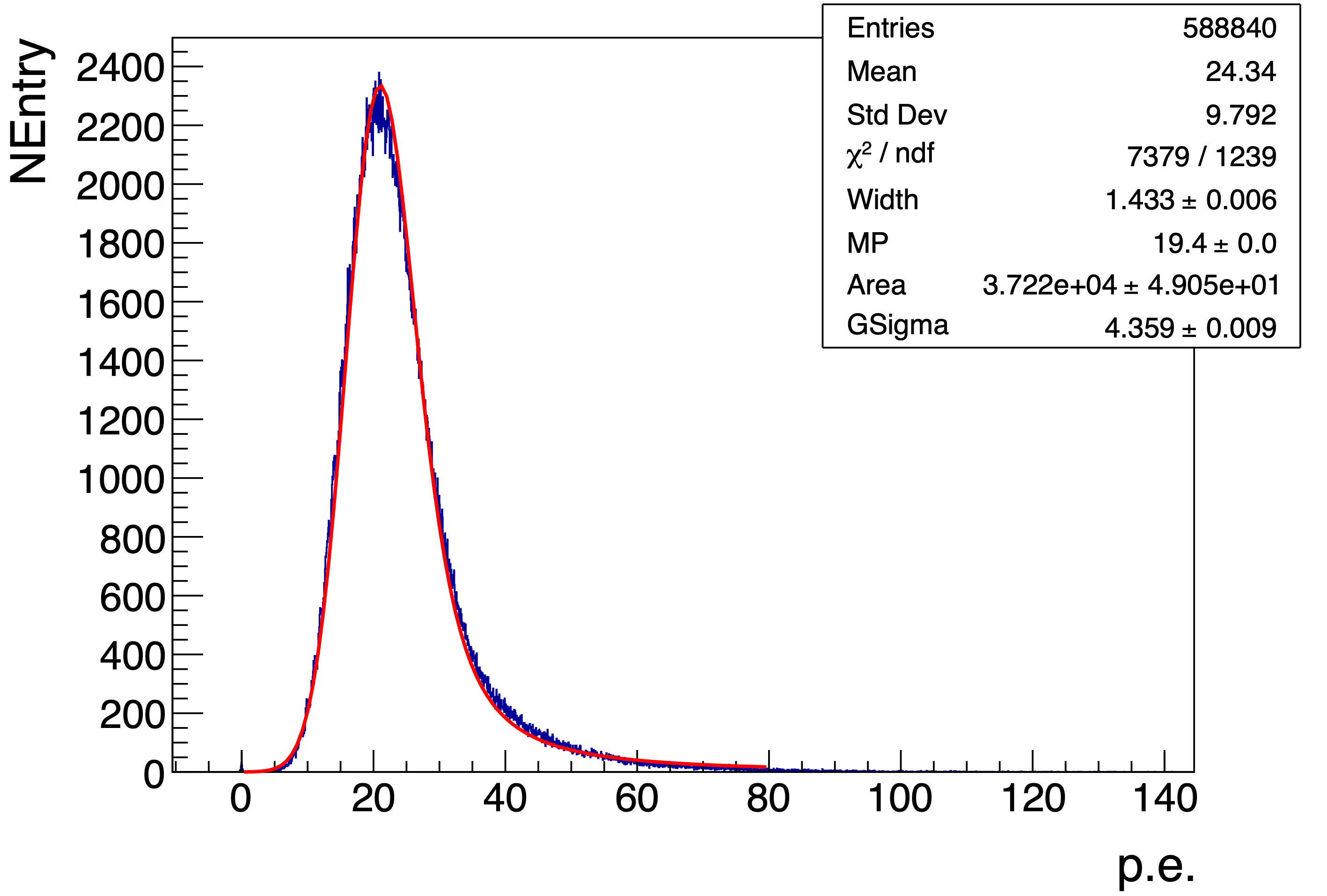} 
    \caption{Y-11}
  \end{subfigure}
  \caption{Light yield distributions for each fiber and the corresponding fit results.}
  \label{lightyield_result}
\end{figure}

The MPPC waveforms are integrated from $-50$~ns to $+100$~ns relative to the pulse peak to obtain a light yield distribution, which is fitted with a Landau-Gaussian convolution function~\cite{rootcode} in the range from 0 to 80~p.e. 
The most probable value (MPV) of the Landau component is taken as the representative light yield.
Figure~\ref{lightyield_result} shows example light yield distributions for each fiber. The BCF-9995XL fiber is excluded due to its low light yield ($\sim$1~p.e.). This is attributed to a significant mismatch between the scintillator's emission spectrum and the fiber's absorption spectrum.

\begin{table}[htb]
    \caption{Measured light yields and uncertainties.}
    \makebox[\textwidth][c]{
    \newcolumntype{C}[1]{>{\centering\arraybackslash}p{#1}}
    \begin{tabular}{l|l||c|c|C{1.5cm}}\hline
        \multicolumn{2}{c||}{} & BCF-92XL & BCF-9929AXL & Y-11 \\ \hline \hline
        \multicolumn{2}{c||}{Light Yield (p.e.)} & 10.54 & 11.50 & 18.01 \\ \hline \hline
        \multirow{4}{*}{Uncertainties}  & Fitting & 0.009 & 0.010 & 0.011 \\ 
        & Reproducibility & 0.650 & 0.614 & 1.014 \\ 
        & Fit Range  & 0.027 & 0.021 & 0.172\\ 
        & Event Selection & 0.146 & 0.187 & 0.291 \\ \hline \hline
        \multicolumn{2}{c||}{Total Uncertainty} & 0.67 & 0.64 & 1.07 \\ \hline
    \end{tabular}
    }
    
    \label{lightyield_result_table}
\end{table}

We consider three sources of systematic uncertainties. The uncertainty in reproducibility arising from sample-to-sample variations (including individual fiber differences, scintillator-to-fiber coupling, and fiber-to-MPPC coupling) is evaluated as the standard deviation of the MPV values measured across the seven fiber samples.
The fit-range dependence is evaluated by varying the start point from 0 to 5~p.e. and the end point from 80 to 110~p.e. The resulting uncertainty is defined as the half the difference between the maximum and minimum results.
The impact of the event selection, which primarily suppresses multi-particle events, is evaluated by repeating the analysis without this selection. The resulting uncertainty is defined as the difference between the MPV values obtained with and without the selection.

The total uncertainty for the light yield measurement is calculated as the quadrature sum of the statistical and systematic uncertainties. The measured light yields, with their statistical and systematic uncertainties, are summarized in Table~\ref{lightyield_result_table}.

The results show that the light yields of BCF-92XL and BCF-9929AXL are approximately 60\% of that of the Y-11 fiber. The BCF-XL fibers are single-clad, whereas Y-11 is multi-clad. The trapping efficiencies of single- and multi-clad Y-11 fibers are reported to be 3.1\% and 5.4\%, respectively~\cite{5}. Since the light yield is expected to scale with the trapping efficiency, the observed light yields of the BCF-XL fibers are comparable to those of a single-clad Y-11 fiber.

\subsection{Time resolution}
The time resolution is evaluated using the same dataset as the light yield measurement. To determine the time resolution, the timing of the MPPC signal is calculated relative to the average timing of the PMT1 and PMT2 signals.

\begin{figure}[htb]
    \centering
    \includegraphics[width=0.7\linewidth]{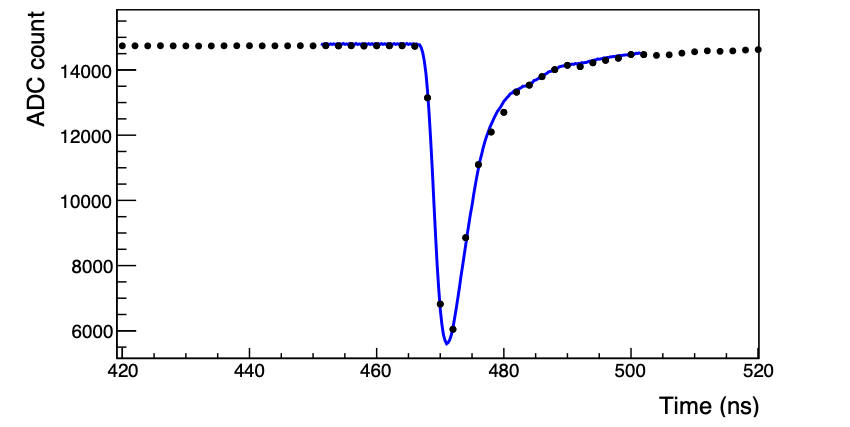}
    \caption{Determination of the timing of PMT waveform using the template (blue). The position of the template shown in the figure is the one obtained by minimizing the sum of squared differences.}
    \label{template}
\end{figure}

For the MPPC signal, the timing is determined using the same linear interpolation method as in the decay time measurement (Section~\ref{decaytime_method}). For the PMT signals, a template-fitting approach is used to accommodate their short pulse and limited sampling points. A template waveform is obtained with a 4 GS/s oscilloscope (wavesurfer 3034z, TELEDYNE LECROY). The PMT waveform is fitted using this template, and the timing is determined by minimizing the sum of squared differences (see Fig.~\ref{template}). To select single incident electron events, the same event selection as in the light yield analysis is applied. The time resolution is then evaluated as the standard deviation of the timing distribution.

\begin{figure}[htb]
  \centering 
  \begin{subfigure}[b]{0.48\textwidth}
    \centering
    \includegraphics[width=\linewidth]{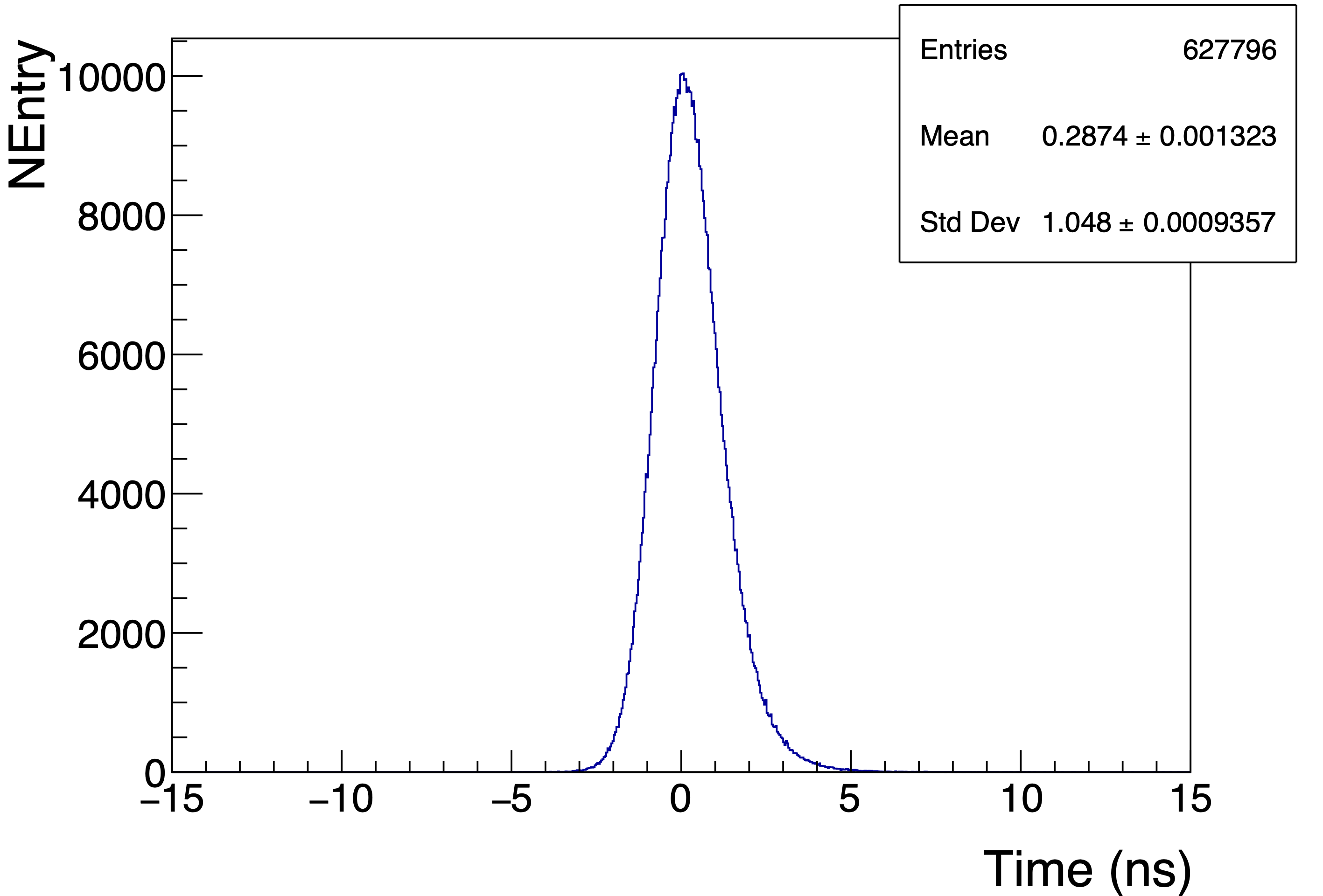}
    \caption{BCF-92XL}
  \end{subfigure}
  \begin{subfigure}[b]{0.48\textwidth}
    \centering
    \includegraphics[width=\linewidth]{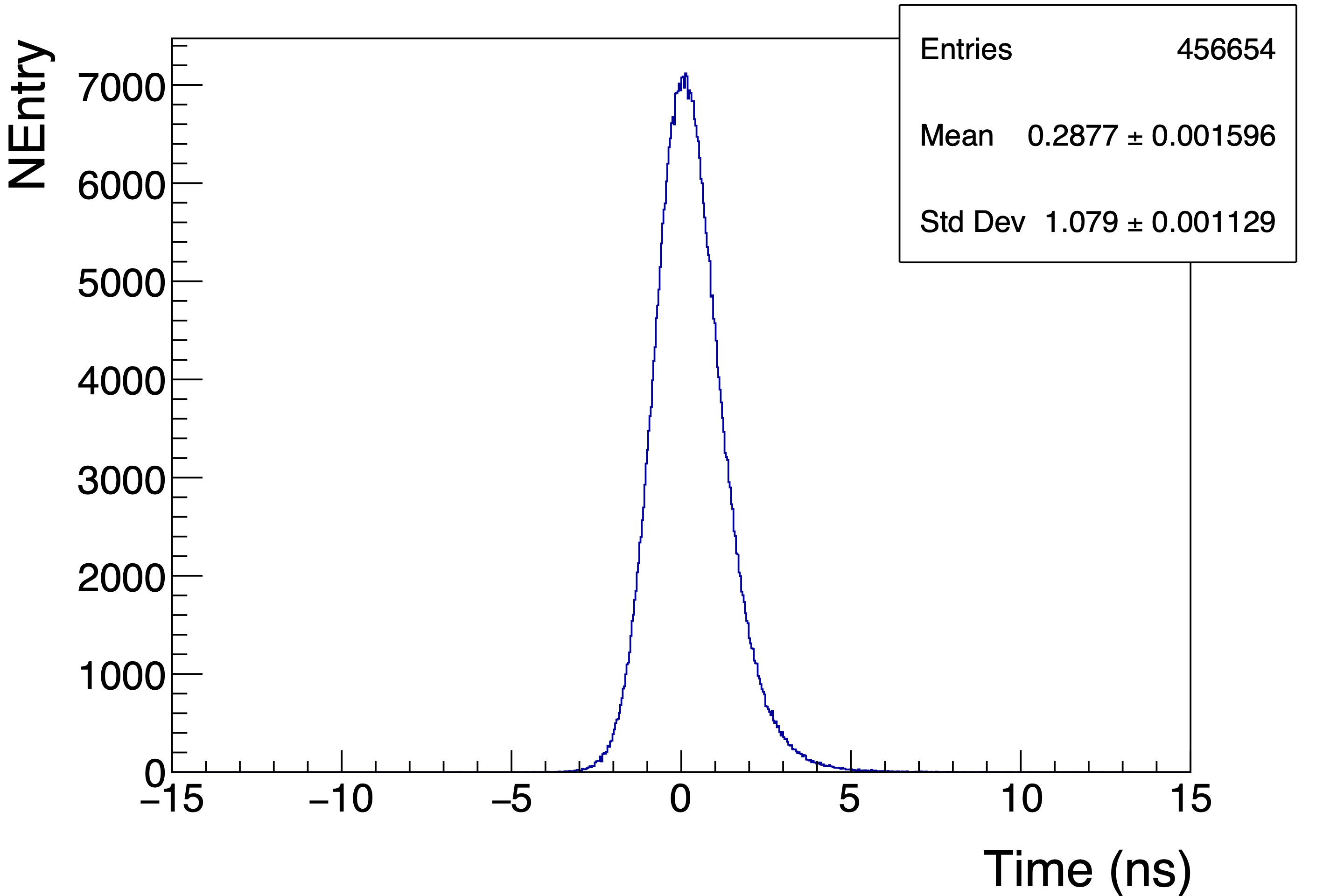} 
    \caption{BCF-9929AXL}
  \end{subfigure}
  \\[1em]
  \begin{subfigure}[b]{0.48\textwidth}
    \centering
    \includegraphics[width=\linewidth]{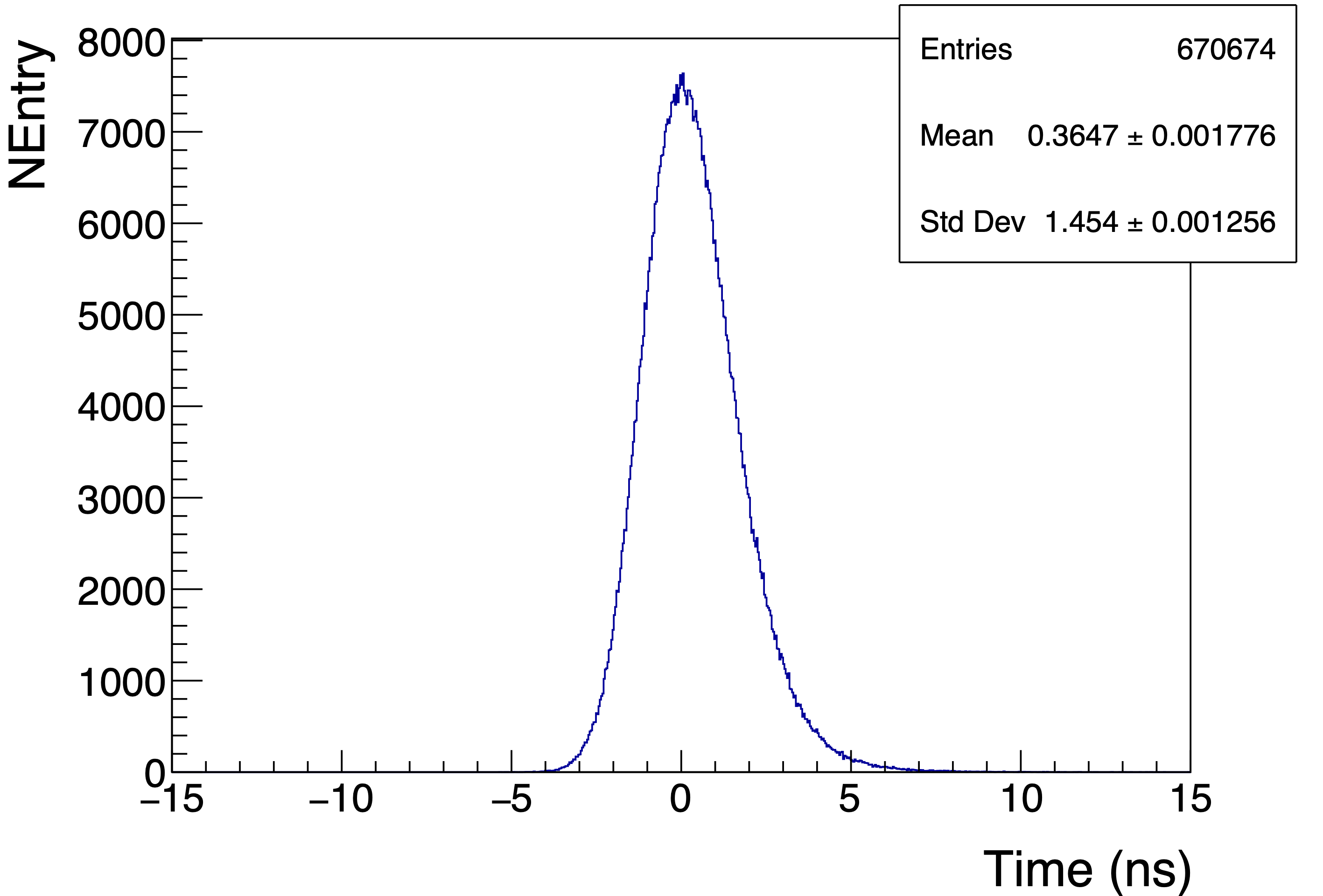}
    \caption{Y-11}
  \end{subfigure}   
  \caption{Timing distributions for each fiber.}
  \label{timeresolution_result}
\end{figure}

Figure~\ref{timeresolution_result} shows example time distributions for the fibers. The average time resolution of seven fibers is taken as the representative time resolution for that fiber, and the standard deviation is taken as the systematic uncertainty arising from sample-to-sample variations.

\begin{table}[htb]
    \caption{Measured time resolutions and uncertainties.}
    \makebox[\textwidth][c]{
    \newcolumntype{C}[1]{>{\centering\arraybackslash}p{#1}}
    \begin{tabular}{l|l||c|c|C{1.5cm}}\hline
        \multicolumn{2}{c||}{} & BCF-92XL & BCF-9929AXL & Y-11  \\ \hline \hline
        \multicolumn{2}{c||}{Time Resolution (ns)} & 1.11 & 1.06 & 1.52 \\ \hline \hline
        \multicolumn{2}{c||}{Uncertainty} & 0.04 & 0.04 & 0.05 \\ \hline
    \end{tabular}
    }
    \label{timeresolution_result_table}
\end{table}

The time-walk effect is corrected, and half the difference between the values before and after the correction is taken as the associated uncertainty. This uncertainty is at most 0.004~ns and is negligible compared to the uncertainty arising from the reproducibility. The results are summarized in Table \ref{timeresolution_result_table}. From these results, the time resolutions of BCF-92XL and BCF-9929AXL are approximately 70\% of that of Y-11.

%% file: src/04_conclusion.tex
\section{Conclusions}
We evaluate the performance of the Luxium BCF-XL series single-clad fibers (BCF-92XL, BCF-9929AXL, and BCF-9995XL) and compare them with the Kuraray Y-11 multi-clad fiber.

In the laser-based setup, we measure the decay time and attenuation length. We confirm that the BCF-XL series have better timing performance than Y-11, with the decay times of approximately 30\% of that of Y-11. The attenuation lengths are found to be comparable to Y-11 within the measurement range of 10 to 320~cm.

Using a 3~GeV/$c$ electron beam, we characterize the practical performance of each fiber coupled to a plastic scintillator EJ-204, focusing on the light yield and time resolution. The light yields are about 60\% of that of Y-11, which is attributed to the difference between single- and multi-clad structures, and are comparable to those of a single-clad Y-11 fiber. Despite the lower light yields, the time resolutions of the BCF-XL series are superior to that of Y-11.